\documentclass{aastex631}

\usepackage{bm}
\usepackage[polish,english]{babel}
\usepackage{polski}
\usepackage[utf8]{inputenc}
\usepackage{amsmath}
\usepackage{booktabs}

\begin{document}

\correspondingauthor{Adam A. Zychowicz}
\email{adam.zychowicz@doctoral.uj.edu.pl}

\title{Bayesian approach to equipartition estimation of magnetic field strength}

\author[0009-0004-2544-0632]{Adam A. Zychowicz}
\affiliation{Astronomical Observatory of the Jagiellonian University, ul. Orla 171, 30-244 Kraków, Poland}
\affiliation{Jagiellonian University, Doctoral School of Exact and Natural Sciences, ul. Prof. St. Łojasiewicza 11, 30-348 Kraków, Poland}

\author[0000-0002-6280-2872]{Krzysztof T. Chyży}
\affiliation{Astronomical Observatory of the Jagiellonian University, ul. Orla 171, 30-244 Kraków, Poland}

\begin{abstract} 
Magnetic fields, together with cosmic rays (CRs), play an important role in the dynamics and evolution of galaxies, but are difficult to estimate. Energy equipartition between magnetic fields and CRs provides a convenient way to approximate magnetic field strength from radio observations. We present a new approach for calculating the equipartition magnetic field strength based on Bayesian methods. In this approach, the magnetic field is a random variable that is distributed according to a posterior distribution conditional on synchrotron emission and the size of the emitting region. It allows the direct application of the general formulas for total and polarized synchrotron radiation without the need to invert these formulas, which has limited the equipartition method to highly simplified cases. We have derived the equipartition condition for the case of different low-energy breaks, slopes, and high-energy cutoffs of power law spectra of the CR proton and electron distributions. The derived formalism was applied in the general case of a magnetic field consisting of both uniform and randomly oriented field components. The applied Bayesian approach naturally provides the uncertainties in the estimated magnetic field strengths resulting from the uncertainties in the observables and the assumed values of the unknown physical parameters. In the examples presented, we used two different Markov Chain Monte Carlo methods to generate the posterior distribution of the magnetic field. We have also developed a web application called BMAG that implements the described approach for different models and observational parameters of real sources.

\end{abstract}



\section{Introduction}
\label{sec:introduction}

Magnetic fields are ubiquitous in the universe, present in the interstellar medium, in clusters of galaxies, and in powerful active galaxies. 
In galaxies such as the Milky Way, the energy contained in magnetic fields generated by dynamo processes is comparable to the turbulent and thermal energy densities found in the interstellar medium \citep[e.g.][]{Beck2015, Han2017}. Consequently, magnetic fields, together with cosmic rays (CRs), play a central role in the dynamics and evolution of galaxies. For example, they can help resist gravitational collapse in molecular clouds and thus influence the process of star formation \citep{Rees1987}. A substantial amount of energy is likely to be released in the phenomenon of interstellar magnetic reconnection, with important implications for particle acceleration and heating of the interstellar medium (ISM) \citep{Wezgowiec2022}. In addition, magnetic fields can trap and guide CRs, confining them and serving as the primary mechanism for their transport. This, in turn, contributes to the initiation and shaping of galactic winds and outflows \citep{Werhahn2023}.

In galaxy clusters, magnetic fields are able to reduce the conductivity in a thermal plasma, as shown by cluster cold fronts \citep[e.g.][]{ZuHone2013}. In cluster radio halos, they tend to be aligned with the expected galaxy merger axis. This is consistent with numerical simulations that predict the existence of turbulent magnetic fields in clusters, stirred and amplified by matter motions \citep{Hu2024}. Magnetic fields also play a fundamental role in understanding how active galactic nuclei (AGNs) operate and influence their host galaxies and the surrounding intergalactic medium. They are crucial for processes such as jet formation and CR acceleration \citep{EHT2021}. According to magnetohydrodynamic cosmological simulations, primordial magnetic seed fields provide additional pressure that affects the evolution of gaseous structures. This suppresses cosmic star formation and alters the mass population of galaxies in the early universe \citep[e.g.][]{Marinacci2016}.

The magnetic field strength values are therefore extremely important for understanding the physics of many astrophysical objects, but are difficult to obtain. It is easier to derive the global geometrical structure of galactic magnetic fields which can be modeled from astrophysical observations, mainly Faraday rotation measurements and the polarized synchrotron emission. Recent developments in infrared observations have also allowed the structure of magnetic fields to be revealed from magnetically aligned dust grains \citep{Borlaff23}. However, these methods provide the orientation of the magnetic field lines, but not the field strength.

Synchrotron radiation theory provides a potential way to estimate the value of the magnetic field strength from the synchrotron intensity but requires additional assumptions due to the unknown value of the number density of radiating CR electrons: $I_{syn} \propto n_e B^{(\gamma+1)/2}$, where $\gamma$ is a power law index in the electron energy distribution $n_e \propto E^{-\gamma}$. This is called synchrotron degeneracy. To eliminate this unknown, the equipartition between the energy of the volume-averaged magnetic field $\epsilon_{B}=B_{eq}^2/(8\pi)$ and the energy in CRs $\epsilon_{cr}$ is usually introduced: $\epsilon_{cr}=\epsilon_{B}$ \citep[e.g.][]{pacholczyk, Govoni2004, Beck2005}. This assumption is similar to a principle of minimizing of the sum of these energies: $\epsilon_{cr} + \epsilon_{B}=min$ \citep{Bell1978a, Bell1978b}. Another important piece of information to know when using these methods is the shape of the proton and electron energy spectra to calculate the CR energy budget. Typically, a power law and a constant ratio of the proton number density to the electron number density $K_0=n_p /n_e$ are assumed. For normal galaxies, the value $K_0 = 100$ is usually used \citep{Beck2005}, while for radio galaxies $K_0 = 1$ or $K_0 = 0$ is often preferred \citep[e.g.][]{Hardcastle1998, Harwood2013}. The value of $K_0=100$ goes back to \citet{Bell1978b}, who found that for CR energy power law spectra with the same index $\gamma$ for electrons and protons, $K_0=(m_p /m_e)^{(\gamma-1)/2}$, where $m_p$, $m_e$ are the proton and electron masses, respectively. This gives $K_0=43$ for $\gamma=2.0$ and $K_0=132$ for $\gamma=2.3$.

In the past, the principle of minimizing energies was often carried out by integrating the emission over a fixed range in radio frequency, usually between 10\,MHz and 100\,GHz \citep{Burbidge1956, pacholczyk, Miley1980}. This assumption is unfortunate because a given frequency of synchrotron observation corresponds to different electron kinetic energies for different magnetic field strengths and therefore to different CR electron number density. Therefore, this criterion was modified to perform minimizing over a fixed interval of CR electron energy \citep{Pohl1993, Pfrommer2004, Beck2005, Arbutina2012, Beck2015} which now depends on the unknown low-momentum cutoff of the CR distribution. A different approach has been taken for clusters, where CR electrons are expected there to be mainly the result of inelastic interactions of CR protons with the thermal gas of the intracluster medium (ICM). This hadronic scenario provides another constraint that eliminates the uncertainty of the electron distribution. In this case, the minimum-energy magnetic field depends only on the energy distribution of the CR ions \citep{Pfrommer2004}. Both hadronic and leptonic scenarios for synchrotron emission have been applied for the starburst galaxies M\,82 and NGC\,253 \citep{Persic2008, Rephaeli2010}. The equipartition magnetic field was obtained there by modeling with a modified GALPROP code developed to study the generation and propagation of CRs in the Milky Way \citep{Strong2007}. The energy equipartition was assumed and implemented iteratively to solve for the energy densities of protons, electrons, and magnetic fields. For example, in NGC\,253 the spectra of low energy CRs are significantly flatter than at high energies, while at higher energies at $E\ge 1$\,GeV the greater energy losses of electrons result in a steeper spectrum than that of protons, so that the electron spectrum is characterized by a power law index $\gamma =2.74$ compared to $\gamma =2.55$ for protons. The assumption that protons and electrons have constant and equal energy spectra indices is not fulfilled in this case.

There is no clear physical justification for the CRs and magnetic fields being close to the energy equipartition or its minimum. It is possible that the existence of the energy equipartition is not necessarily determined by causally related direct processes between CRs and magnetic fields but may result from the interplay of various processes already observed in galaxies or active objects. For example, the escape of CRs in front of the shock is an important part of the overall particle acceleration process. The CRs drive the amplification of the magnetic field, which in turn governs the number of escaping CRs. If a smaller number of CRs escape, the magnetic field would not be sufficiently amplified to confine and accelerate the CRs. If a larger number of CRs had escaped, the magnetic field would have increased too quickly to allow them to escape. Therefore, a self-regulating system is organized to allow an appropriate number of CRs to escape \citep{Bell2013}. In galaxies, the magnetic field can be stretched and tangled by turbulent motions in the ISM which are regulated by star formation activity \citep{Schleicher2016}. A small-scale dynamo exponentially amplifies the magnetic field until it is approximately equal to the turbulent energy. According to magnetohydrodynamic (MHD) galaxy simulations \citep{Pfrommer2022}, in the Milky Way-mass galaxies after saturation at small scales, the magnetic fields continue to grow at larger scales to reach equipartition with thermal and CR energies. \citet{Seta2019} argued that equipartition is valid for normal star-forming spiral galaxies at scales above about 1\,kpc, but probably not at smaller scales and not for galaxies undergoing a massive starburst. In starbursts, high gas densities and mass loss due to winds can lead to significant losses of CR energy density. Consequently, the magnetic fields must be much stronger for the synchrotron emission to match the observed radio brightness. Magnetic fields may also be strongly spatially correlated with sites of significant star formation. In such situations, the magnetic energy density may generally be out of equipartition and significantly exceed the CR energy density, especially in central molecular zones \citep{Yoast-Hull2016}.
In addition, CR diffusive transport leads to lower CR density at CR electron acceleration sites (in star-forming regions) and higher CR density away from them, which may result in underestimation of equipartition magnetic fields in star-forming regions and overestimation of the fields away from these regions.

In the solar neighborhood, typical values of the energy densities of CRs, magnetic fields, radiation fields, and gas kinetic energy are comparable and on the order of 1\,eV\,cm$^{-3}$ \citep[e.g.][]{Badhwar1977, Boulares1990}. Arguments for the validity of the equipartition on large (kpc) scales come from the joint analysis of radio continuum and $\gamma$-ray data, which allowed an independent determination of the total magnetic field strengths in the Milky Way of 6\,$\mu$G \citep{Strong2000}. The same value was derived from energy equipartition \citep{Berkhuijsen2001}. 
The direct measurements of magnetic fields in the local ISM by
Voyager 1 and Voyager 2 of $4.8 \pm 0.4\,\mu$G and $6.8 \pm 0.3\,\mu$G, respectively, are in excellent agreement with the equipartition value.

The study of magnetic fields in the Large Magellanic Cloud (LMC) showed agreement between the equipartition magnetic field value and estimates derived from Faraday rotation measurements of extragalactic polarized sources behind the LMC \citep{Mao2012}. Furthermore, this value is consistent with the upper limit of the magnetic field strength inferred independently of the equipartition hypothesis, using the galaxy's observed gamma-ray flux. This result contradicts the previous claim of \citep{Chi1993} that the equipartition assumption is violated in the LMC. In another galaxy, the nearby spiral M\,51, the ordered magnetic field derived from the observed polarized emission and the equipartition assumption was 4 times larger than estimates from the Faraday rotation, but this difference can be well explained by the contribution of an anisotropic random magnetic field to the polarized signal \citep{Fletcher2011}. 

Although the principle of equipartition or minimum energy has been widely used in various works, we should keep in mind only the approximate values they give and the limitations they have. Equipartition between CRs and magnetic fields is unlikely to be valid on small spatial scales, smaller than the diffusion length of CRs, and on short time scales, shorter than the diffusion time of CRs. 
A constant value of $K_0=100$ often used for galaxies is motivated by the same momentum injection spectrum of primary CR electrons and protons from supernovae by diffusive shock acceleration and by the measurements in the Milky Way \citep{Bell1978a, Bell2004}. This assumption may hold close to CR injection sites, but not across galaxies \citep{Ponnada2024}. Actual energy spectra result from injection processes, CR transport, and energy losses which are different for protons and electrons. 
For example, at energies below a few hundred MeV, electrons and protons lose energy mainly through Coulomb interactions with gas particles, leading to ionization of neutral or charged ions and electronic excitations \citep[e.g.][and references therein]{Ruszkowski2023}. Electrons can also radiatively dissipate energy in the bremsstrahlung process. For electrons at higher energies, the dominant electron energy losses are via synchrotron emission and inverse Compton (IC) scattering at radiation fields. Protons colliding inelastically with the surrounding gas lose energy in the production of gamma rays via pion $\pi^0$ and in production of secondary cosmic ray leptons via $\pi^{+-}$ decay. This can violate the equipartition principle if these processes are not explicitly modeled in the energy balance calculations. For example, using the constant value of $K_0$ underestimates the total magnetic field in the outer disks and halos of galaxies, where the emitting CR electrons propagate far away from the sites of their acceleration and where the energy losses are significant \citep{Heesen2023}. 

In dense gas, e.g. in starburst regions, synchrotron emission may be dominated by secondary electrons and positrons from hadronic processes \citep{Lacki2013} which should be included in modeling. The one-zone model of CR electrons (primary and secondary), secondary positrons, and protons in the starburst galaxy M\,82, using multifrequency radio and $\gamma$-ray data, yielding a magnetic field of 275\,$\mu$G \citep{Yoast-Hull2013}. This value is within a factor of two of similar modeling but assuming equipartition \citep[150\,$\mu$G,][]{deCea2009}, indicating discrepancies to be reckoned with.

The synchrotron degeneracy in determining the magnetic field strength and other properties of radio sources could be overcome by additional measurements of the X–rays or $\gamma$-rays. These high-energy photons may be produced by: primary or secondary population of leptons in the IC process on the cosmic microwave background, synchrotron photons or other local photon fields, bremsstrahlung emission of relativistic electrons, and by protons through pion decay. Depending on the available data and accepted processes, the population of protons and electrons could be constrained, and with measured synchrotron radio emission the magnetic field can be obtained directly without relying on the equipartition assumption. As mentioned above, such methods have been applied to the Milky Way, some starburst galaxies, and clusters of galaxies. A similar approach was applied to the study of X-ray emission from the lobes of 33 classical double radio galaxies and quasars, where magnetic field strengths were found to be 35\% of the equipartition value \citep{Croston2005}. For these objects, the equipartition was assumed to be only between the radiating electrons and the magnetic fields. Some other radio galaxies \citep[e.g.][]{Brunetti97, Brunetti2002, Hardcastle2002, Konar2009} were also found to support the equipartition assumption for the lobes or hotspots within the factor $\approx 2-3$. This required modeling of the X-ray spectrum, separation of the spatial components, and a good constraint on the origin of the X-ray emission. For many galaxies and radio galaxies, especially the more distant ones, the observational data are insufficient for such detailed modeling and accurate detection of the processes shaping the radio and high-energy emissions. In these situations, the equipartition method remains a useful tool for obtaining an approximate insight into the magnetic field properties in these objects.

The new era of radio astronomy instruments, such as the Square Kilometre Array (SKA) or the LOw Frequency ARray (LOFAR), requires new research methods capable of exploiting the extensive knowledge of radio spectra in various objects, capable of removing previous limitations. In this paper, we present a new approach to calculating the magnetic field strength based on the principle of energy equipartition. Our approach uses Bayesian statistics to estimate the magnetic field as a posterior distribution. Traditional methods using energy equipartition, as seen in the work of \citep{Miley1980, Beck2005}, derive the magnetic field strength from the observed synchrotron emission, thus solving the inverse problem of synchrotron emission theory, which provides synchrotron radiation from a known magnetic field strength (direct problem). The search for an analytical formula for the solved inverse problem has so far severely limited its applicability, mostly reducing this method to cases with only ordered or random magnetic fields, a fixed ratio of proton to electron number densities and simplified power law forms of the CRs energy spectra. It has not been possible to apply it to situations with significant energy losses due to processes such as synchrotron cooling or scattering due to the IC effect, which lead to curved energy spectra and affect protons and electrons differently. By using the Bayesian method, we avoid the need to solve the inverse problem directly. This approach allows the posterior distribution of the magnetic field to be derived using direct formulas for synchrotron radiation from a mixture of uniform and random components. Another advantage of using the Bayesian method is that it inherently provides a framework for generating realistic confidence intervals for the estimated magnetic field values. It also facilitates the evaluation of the effects of model assumptions and uncertainties in the observational parameters. 

We use Markov Chain Monte Carlo (MCMC) methods to generate the posterior distribution of the magnetic field strength. In the example calculations, we use two independent simulation codes to ensure that our results are not biased by the computational approach. We derive the equipartition condition for the case of different low-energy breaks of the proton and electron particle distributions and different slopes of their power law spectra. This example illustrates the ability of the Bayesian method to easily handle scenarios where the ratio of proton to electron number densities is not constant. We also recommend how this method can be used and how it can be further developed. Moreover, we have developed a web application that applies this approach under real astrophysical conditions, making our method adaptable to different values of model and observational parameters.

The structure of our paper is as follows: in the next section, we present how the Bayesian approach can be used to solve inverse problems in general. In Section~\ref{sec:equip} we present the proposed description of the geometric configuration of the magnetic field, the energy spectra of the CRs, the corresponding synchrotron emission formula, and the equipartition principle. We also present a theoretical formula for the equipartition magnetic field in the case of purely uniform or random fields. In Section~\ref{sec:bayesian} we formulate a Bayesian approach to the equipartition problem and in Section~\ref{sec:example} we present an example of MCMC simulations. The results are discussed in Section~\ref{sec:discussion}, which also reports on the web application and provides recommendations for using the method for galaxies and radio galaxies. Finally, we summarize our results in Section~\ref{sec:summary}.

\section{Bayesian method for inverse problem}

The problem of estimating the magnetic field strength under the assumption of an energy equipartition involves a combination of many physical parameters, which are either observed with measurement uncertainties, or simply not directly known. For this reason, we will use the Bayesian method to compute the probability distribution of the magnetic field taking a specific value, which will provide a more complete understanding of the field strength given our limited knowledge of physical parameters and noisy measurements. We apply the Bayes theorem to compute the posterior $\pi_{\text{posterior}}$ probability distribution \citep[e.g.:][]{tutorial_on_Bayesian}:
\begin{equation}
\label{eq:bayesian_general}
\pi_{\text{posterior}}(\bm{\theta} | \bm{D},\bm{M}) = \frac{\pi_{\text{likelihood}}(\bm{D} | \bm{\theta},\bm{M}) \, \pi_{\text{prior}}(\bm{\theta})} {\int \pi_{\text{likelihood}}(\bm{D} | \bm{\theta},\bm{M}) \, \pi_{\text{prior}}(\bm{\theta}) d\bm{\theta}}=\frac{\pi_{\text{likelihood}}(\bm{D} | \bm{\theta},\bm{M}) \, \pi_{\text{prior}}(\bm{\theta})} { \pi_{\text{evidence}}(\bm{D}|\bm{M})}, 
\end{equation}
where $\bm{\theta}$ is the vector of parameters, $\bm{D}$ is a vector of data, usually some measured quantities that we can model, and $\bm{M}$ are remaining model parameters of fixed values that we assume to be known. It is important to remember that each model requires a considerable number of parameters $\bm{M}$ to be fixed, but the dependence of the distributions on these parameters is omitted in the formula shown in the following sections. The posterior distribution describes the probability that the parameters $\bm{\theta}$ take a specific value, given the measured data $\bm{D}$. The likelihood function $\pi_{\text{likelihood}}$ describes the probability of measuring values $\bm{D}$, given a set of parameters $\bm{\theta}$. The prior distribution $\pi_{\text{prior}}$ encodes any prior knowledge that we have about the physically acceptable ranges or parameters from any source, including, but not limited to: previous measurements, other theoretical considerations, or even an educated guesses. The quantity $\pi_{\text{evidence}}$ in the denominator is the Bayesian evidence, which is a normalization constant that is independent of the model parameters $\bm{\theta}$ and can be neglected in parameter inference. 

The use of Bayesian reasoning to solve inverse problems is a procedure which can be applied to numerous problems not only in astronomy but in various branches of experimental and observational sciences to problems of wide range of complexity (see \citet{Tarantola, tutorial_on_Bayesian} for some examples). Without use of Bayesian method, to solve inverse problem, one would need to model observed quantity as a function of parameters of interest, and then try to invert resulting formulas hoping to recover sought after parameters. If this direct solution of inverse problem is possible then it usually results in convenient closed-form formula allowing for direct computation of parameters. Even then, the propagation of uncertainties of observed quantities to resulting values of parameters needs to be considered. Unfortunately, for systems complex enough to model, the resulting formula might be impossible to be inverted in general case, as is the case with the synchrotron intensity formulas derived in this work. Equation (\ref{eq:bayesian_general}) in turn describes distribution of unknown parameters of interest \bm{$\theta$} (in our case magnetic field strength) as a product of prior distribution with likelihood which depends on observed synchrotron emission intensity and modeled synchrotron intensity as a function of magnetic field strength and geometry. Synchrotron intensity for presupposed magnetic field configuration can be derived from the theory of synchrotron emission. The use of Bayesian methods for inverse problem means that it is not necessary to directly invert the synchrotron emission formulas and in turn, opens the possibility of including more elaborate choices of magnetic field configuration and CR energy spectrum, which would lead to otherwise not invertible solutions. Bayesian approach will lead to determination of magnetic field strength not as a single value, but as the posterior distribution of the values of modeled magnetic field strength, which additionally provides a natural way of defining uncertainties on the derived values. In Section~\ref{sec:equip} we derive formulas for total and polarized intensity for synchrotron radiation for a given magnetic field geometry (forward problem), and in Section~\ref{sec:bayesian} we discuss how to construct quantities present in Equation (\ref{eq:bayesian_general}), and thus how to apply the resulting formulas in the Bayesian formulation of the inverse problem.

\section{New approach to equipartition}
\label{sec:equip}

\subsection{Synchrotron emission formulas }
\label{ssec:synch_em_formulas}
\setcounter{footnote}{2}

In order to solve the inverse problem, we need formulas for the synchrotron intensity under the energy equipartition. Synchrotron emission occurs when a relativistic electron is accelerated by the presence of a magnetic field. This electron of charge $e$ and energy $E$ in a uniform magnetic field of strength $B$ spirals along the field and radiates, with a peak of the emission around the characteristic frequency\footnote{All formulas and dimensions in this paper use the cgs unit system.}:
\begin{equation}
    \nu_c = \frac{3e B \sin \alpha}{4 \pi mc} \left(\frac{E}{mc^2}\right) ^2 =  1.608 \times 10^{-2} \,  {\rm{GHz}} \left(\frac{B \sin \alpha} {\mu \rm{G}} \right) \left(\frac{E}{\rm{GeV}}\right)^2
\end{equation}
where $\alpha$ is the electron pitch angle (the angle between the electron velocity and the field).  
We consider the synchrotron emission from an ensemble of electrons in a volume $V$ with density following a power law spectrum with an index $\gamma_e$: 
\begin{equation}
\label{eq:powerlaw}
    n_e(E)= N_e E^{-\gamma_e}
\end{equation}
where $N_e$ is a constant.
We assume that in the observed frequency range the emission is from electrons of this spectral shape, and emission from outside the energy range corresponding to this spectrum is negligible. The general formulas for that emission in the magnetic field of arbitrary geometry have been derived, e.g. by \citet{KS62}. We have rewritten these formulas, using definitions of constants and naming conventions consistent with more recent work to facilitate later comparison of results. The Stokes parameters of the total and linearly polarized flux density (power per unit area and unit frequency range) are in this case as follows:

\begin{align}
\label{eq:KS18_beckconst} 
\begin{split}
\mathcal{I}&=c_2 \left(\frac{2c_1}{\nu}\right)^{\frac{\gamma_e-1}{2}}\frac{N_e}{r^2} \int_V \left( B \sin{\mu} \right) ^{(\gamma_e+1)/2}dV,\\
\mathcal{Q}&= p_0 c_2 \left(\frac{2c_1}{\nu}\right)^{\frac{\gamma_e-1}{2}}\frac{N_e}{r^2} \int_V \left( B \sin{\mu} \right) ^{(\gamma_e+1)/2} \cos{2\chi}dV,\\
\mathcal{U}&= p_0 c_2 \left(\frac{2c_1}{\nu}\right)^{\frac{\gamma_e-1}{2}}\frac{N_e}{r^2} \int_V \left( B \sin{\mu} \right) ^{(\gamma_e+1)/2} \sin{2\chi}dV,
\end{split}
\end{align}

\noindent where the symbols used are:
\begin{align}
\begin{split}
c_1&=\frac{3e}{4\pi m^3 c^5},\\
c_2&=\frac{1}{p_0}\frac{c_3}{4} \Gamma\left(\frac{3\gamma_e-1}{12}\right) \Gamma\left(\frac{3\gamma_e+7}{12}\right),\\
c_3&=\frac{\sqrt{3} e^3}{4\pi m_e c^2},\\
p_0&=\frac{\gamma_e+1}{\gamma_e+\frac{7}{3}},
\end{split}
\end{align}
$r$ is the distance from the radiating area. The angles $\mu$ and $\chi$ describe the orientation of the magnetic field vector $\bm{B}$ in a spherical coordinate system with the pole aligned with the line of sight of the observer. This means that $\mu$ is the angle between the magnetic field and the viewing angle, while $\chi$ is the angle between the sky component of the magnetic field and an arbitrary direction in the sky plane. The use of a power law energy spectrum of electrons (Equation \ref{eq:powerlaw}) gives a power law synchrotron spectrum (Equation \ref{eq:KS18_beckconst}) with a spectral index $\alpha=(\gamma_e - 1)/2$ according to the convention $\mathcal{I}\propto \nu^{-\alpha}$.

\subsection{Magnetic field configuration}
\label{sec:field_configuration}

Since it is rather difficult to obtain general statements valid for all possible magnetic field configurations and different relativistic electron distributions, only limiting cases have often been discussed. Typically, a homogeneous magnetic field with a uniform direction or a field with a completely random orientation has been considered \citep{pacholczyk, Longair1994, Brunetti97, tools,  Beck2005} or a constraint on the CR electrons energy spectral index has been applied \citep[e.g., $\gamma_e=3$;][]{,Sokoloff1998}. Although such approaches may be adequate for conditions in certain regions of both normal galaxies and AGNs, consideration of the proper components of the magnetic field may provide the basis for further important analyses. For example \citet{Chyzy2008_NGC4254SFR} presented magnetic maps of the galaxy NGC\,4254 which allowed the analysis of the relationship of the extracted magnetic field components with the galaxy's location and local star formation activity \citep[a similar study was performed for the galaxy NGC\,6946 by][]{Tabatabaei2013}. In general, the turbulent component of the field may be related to the thermal energy of the gas in galaxies or the action of a small-scale dynamo, while the homogeneous component may be the result of the strong action of a large-scale dynamo \citep{Chyzy_what_drives}. Stretching and shearing motions of the plasma can change the topology of the generated magnetic field and lead to its ordering by galactic winds or tidal interactions without the action of a large-scale dynamo \citep{Vollmer2010, Drzazga2011}. Similar processes may occur in AGNs, where jets are the natural culprit for the flow of magnetized plasma. Therefore, when determining the value of the magnetic field strength, it is necessary to allow for the description of a field with both ordered and turbulent properties. In some works, an approximation of \citet{Segalovitz76} has been used together with the equipartition assumption, where the magnetic field was separated into homogeneous and turbulent components occupying spatially distinct regions of the volume under consideration \citep[e.g.][]{Tabatabaei2008}. In our work we take a different approach. In the absence of observables that would allow us to directly estimate the properties of the random field, in the following, we use a simplified method to describe it. 

We consider the magnetic field configuration according to \citet{KS62}, which consists of two vector components: $\bm{B_u}$ -- ``uniform'' component which is the idealized regular field of constant magnitude and constant direction in the considered volume element;  $\bm{B_r}$ -- isotropic random component of constant magnitude but random direction, representing, e.g. an isotropic turbulent field. The total magnetic field (pseudo) vector is then the vector sum of the uniform and isotropic random field components: $\bm{B}=\bm{B_u} + \bm{B_r}$. Both components contribute to the total synchrotron intensity (Stokes $\mathcal{I}$) but $\bm{B_u}$ is required to give rise to linear polarization (Stokes $\mathcal{Q}$ and $\mathcal{U}$). Synchrotron radiation from a perfectly isotropic random magnetic field is unpolarized (Section \ref{sec:purely_random}). Since the synchrotron linear polarization depends only on the orientation but not on the direction of the magnetic field, the observed polarized radiation can be produced by a regular field with a given direction, a regular field with an opposite direction, a field with multiple field reversals, or an anisotropic random field with a preferred orientation along one spatial coordinate. Determining which type of field geometry dominates the observed polarization requires either additional observations (such as Faraday rotation measurements) or additional knowledge of the source origin.

The angles $\mu$ and $\chi$ describing the orientation in spherical coordinates of the total magnetic field vector $\bm{B}$ can be expressed in terms of the angles $\phi$ and $\theta$ describing the orientation of the random vector component \citep{KS62}:

\begin{align}
\begin{split}
    \label{eq:sin}
    B\sin{\mu}&=(B_u^2\sin^2{\theta_0}+2 B_r B_u \sin{\theta_0}\sin{\theta}\sin{\phi}+B_r^2\sin^2{\theta})^{1/2}, \\
    \cos{2\chi}&=\frac{1}{(B\sin{\mu})^2}(B_u^2 \sin^2{\theta_0}+2B_r B_u \sin{\theta_0}\sin{\theta}\sin{\phi}-B_r^2 \sin^2{\theta}\cos{2\phi}).
\end{split}
\end{align}
The angle $\theta_0$ describes the orientation of the uniform component: if $\theta_0=0$, then $\bm{B_u}$ is parallel to the line of sight and for $\theta_0=\pi/2$ the vector $\bm{B_u}$ lies in the plane of the sky. This choice of coordinate system $(\phi, \theta)$ is made so that $\sin{2\chi}$ averages to zero, giving $\mathcal{U}=0$. In this case, the total linear polarization is described only by $\mathcal{Q}$, now denoted by $\mathcal{P}$ to emphasize that it describes the total linearly polarized emission. In the case of our assumed isotropic distribution of the directions of the random field component, we can replace the integration over the emitting volume by averaging over all possible orientations of the random field, and then multiplying by the volume of the emitting region: $\int_V dV \xrightarrow{} \frac{f V}{4\pi}\int_{4\pi} d\Omega $ \citep{KS62}. We have also introduced $f$ -- the fraction of the volume filled by particles and magnetic fields (the so-called filling factor). After such treatments, Equations (\ref{eq:KS18_beckconst}) take the form:

\begin{align}
\label{eq:KS18_dOmega} 
\begin{split}
\mathcal{I}&=c_2 \left(\frac{2c_1}{\nu}\right)^{\frac{\gamma_e-1}{2}}\frac{f V}{r^2} \frac{N_e}{4 \pi} \int_{4\pi} (B \sin{\mu})^{(\gamma_e+1)/2}d\Omega,\\
\mathcal{P}&= p_0 c_2 \left(\frac{2c_1}{\nu}\right)^{\frac{\gamma_e-1}{2}}\frac{f V}{r^2} \frac{N_e}{4 \pi} \int_{4\pi} (B\sin{\mu})^{(\gamma_e+1)/2} \cos{2\chi}d\Omega,
\end{split}
\end{align}

\noindent where $d\Omega=\sin{\theta}d\theta d\phi$ and the term $B\sin{\mu}$ is related to $(\theta, \phi)$ by Equations (\ref{eq:sin}). The factor $V/r^2$ depends on the geometry of the emitting volume; here we consider a cylinder with base area $A$ and length $l$. This area corresponds to the size of the telescope beam solid angle $\Omega_b$ or the source solid angle. Therefore:
\begin{equation}
    \frac{V}{r^2}=\frac{Al}{r^2}=\frac{\Omega_b r^2l}{r^2}=\Omega_b l.
\end{equation}
For a Gaussian beam with FWHM major axis $\theta_{maj}$ and minor axis $\theta_{min}$, the telescope beam solid angle can be calculated as $\Omega_{b}=\frac{\pi}{4\ln 2} \theta_{maj}\theta_{min}$. To obtain the expression for the specific intensity (the power per unit area, per frequency, and beam solid angle) we divide the expressions for $\mathcal{I}$ and $\mathcal{P}$ by $\Omega_b$, giving expressions for the total specific intensity $I_\nu$ and the linearly polarized intensity $PI_\nu$:

\begin{align}
\label{eq:KS18_intensity} 
\begin{split}
I_\nu=\frac{\mathcal{I}}{\Omega_b}&=c_2 \left(\frac{2c_1}{\nu}\right)^{\frac{\gamma_e-1}{2}}  \frac{f l N_e}{4 \pi} \int_{4\pi} (B \sin{\mu})^{(\gamma_e+1)/2}d\Omega,\\
{PI}_\nu=\frac{\mathcal{P}}{\Omega_b}&= p_0 c_2 \left(\frac{2c_1}{\nu}\right)^{\frac{\gamma_e-1}{2}}  \frac{f l N_e}{4 \pi} \int_{4\pi} (B\sin{\mu})^{(\gamma_e+1)/2} \cos{2\chi}d\Omega.
\end{split}
\end{align}

These equations have analytical solutions only for the two cases where $\gamma_e =3$ or $\gamma_e=7$, and for the two limiting approximations when only uniform or random fields are considered. In all other cases they must be solved numerically.

\subsection{Energy equipartition}

\subsubsection{CR energy spectra}
\label{sec:energy_spectrum}

\begin{figure}[h!]
\begin{center}
\includegraphics[width=5.9cm]{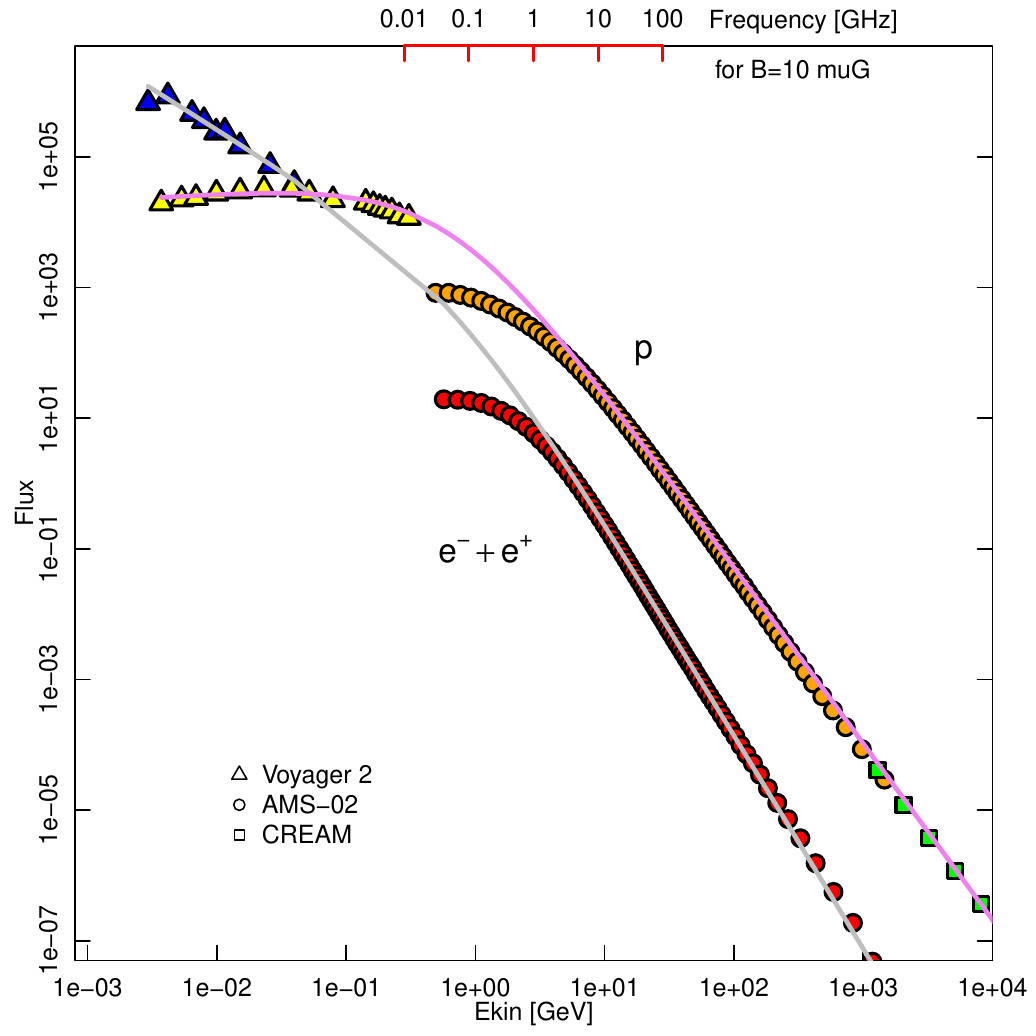}
\includegraphics[width=5.9cm]{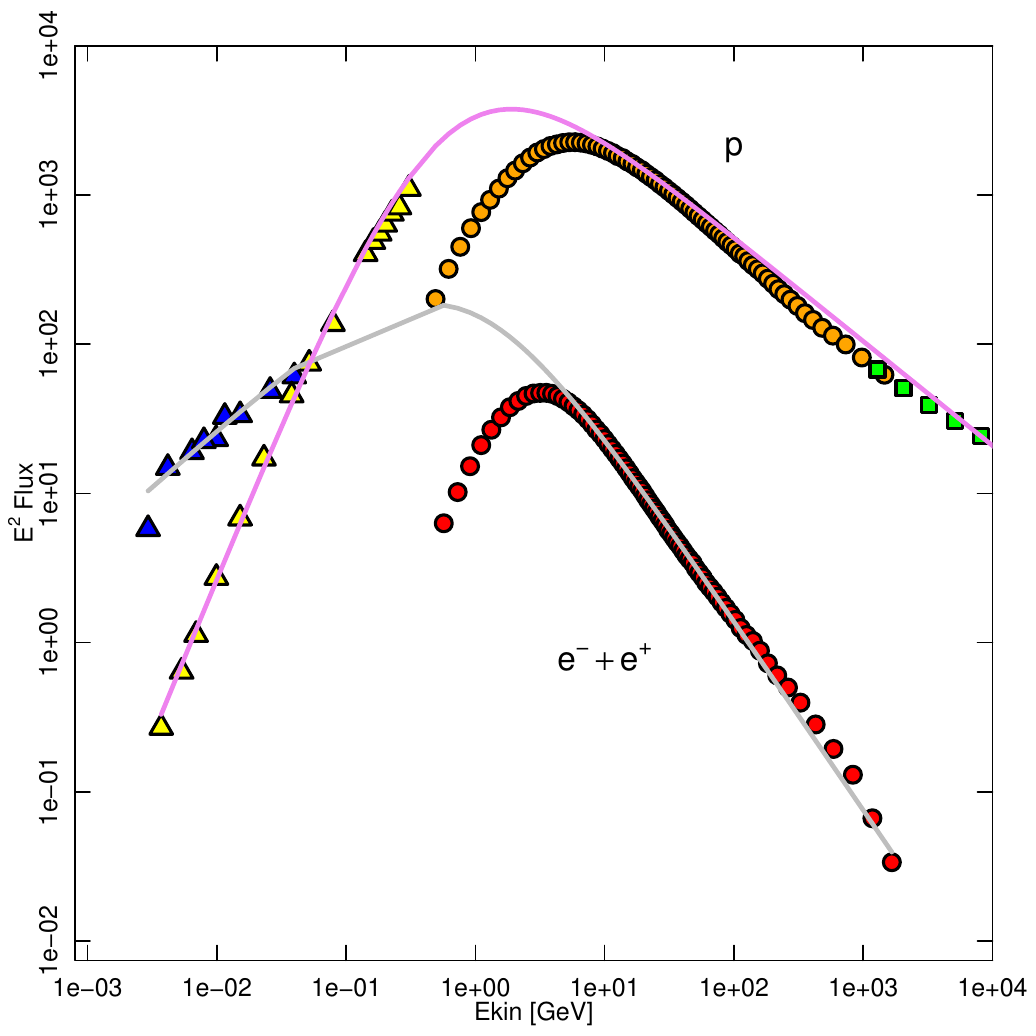}
\includegraphics[width=5.9cm]{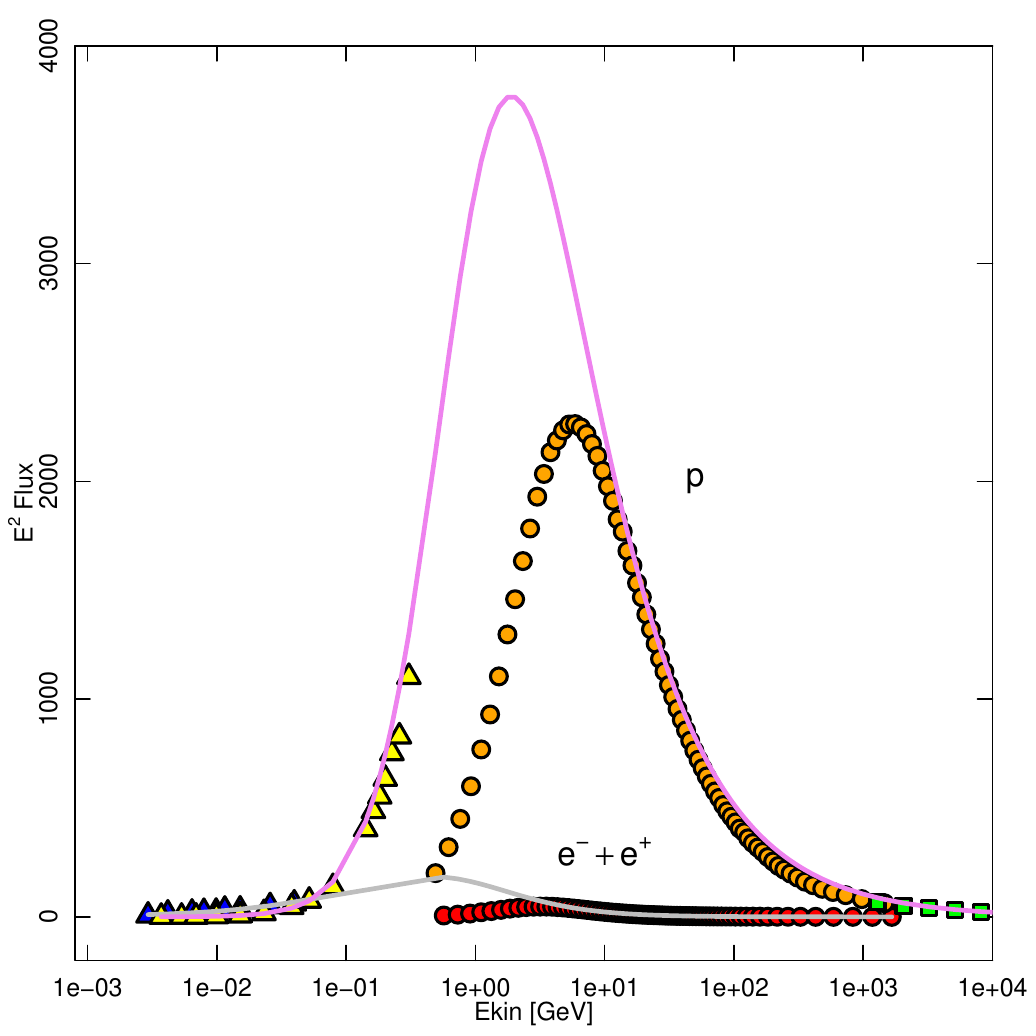}
\caption{Local Galactic differential energy spectra of the CR flux (in particles per $(\rm{m}^{2}\, \rm{s\, sr \, GeV})$) as a function of kinetic energy (in GeV) for protons (p), and electrons together with positrons ($\rm{e}^- + \rm{e}^+$). Data were taken from Voyager 2 after crossing the heliopause (shown as triangles), the International Space Station experiment AMS-02 (circles), and the balloon experiment CREAM (squares). b) The spectra as in a) but scaled by the square of the particle kinetic energy, giving a quantity proportional to the CR energy density per energy decade. c) The same as in b) but without the logarithmic energy scale. The solid lines represent the calculated CR proton (magenta) and electron with positron (gray) spectra according to the broken power law models of \citet{Phan2018}. The AMS-02 spectra below 10\,GeV are usually interpreted as influenced (modulated) by the solar system.}
\label{fig:energy}
\end{center}
\end{figure}

Equations (\ref{eq:KS18_intensity}) describe emission from the power law part of the cosmic ray energy spectrum of electrons. The determination of the equipartition magnetic field strength also requires a quantification of the total energy contained in the CRs. For that, information on the entire energy spectrum of CR protons and electrons is needed. In star-forming galaxies, the main acceleration of particles is likely to occur in strong shocks at supernova remnants, often referred to as diffusive shock acceleration (DSA). In this process, the acceleration is first-order in the shock velocity and automatically leads to a power law spectrum with a spectral energy index $\gamma \approx 2$ for energies greater than the rest energy of the particles and a flatter spectrum for lower energies \citep{Blandford1978, Bell1978a, Bell1978b, Bell2013}. 
The spectral index $\gamma=2.0$ is achieved only for very
strong shocks of an infinite Mach number, because the Mach number $M$ and the spectral index are related as: $\gamma = 1 + (M^2+1)/(M^2-1)$. Weaker shocks with e.g. $M=2$ lead to steeper spectra with $\gamma=2.67$.
However, the energy spectrum of CRs escaping into the surrounding plasma need not be the same as that in the shock during acceleration. Once CRs are accelerated, their propagation through the surrounding medium can change their energy distribution through processes such as energy-dependent diffusion or CR streaming. 

The shape of the CR spectrum can be further modified by various energy losses as the cosmic rays pass through the interstellar medium. In particular, high-energy protons can lose energy through hadronic interactions with protons of the surrounding gas, leading to the production of $\gamma$-rays, secondary electrons/positrons, and neutrinos \citep[e.g.][]{Ruszkowski2023}. On the other hand, Coulomb interactions, including ionization, lead to the removal of the low-energy protons to the ambient plasma. Hence, the transition from low energy ionization/Coulomb energy losses to pion production and adiabatic deceleration losses at higher energies can shape the low energy break in the proton distribution \citep{Schlickeiser2014}. Ionization losses are also important for low-energy CR electrons. In addition, they are sensitive to IC and synchrotron losses, which reduce the number of high-energy electrons and steepen the high-energy electron spectra. Therefore, the actual distribution of CRs in galaxies depends on the history of CR acceleration and interaction with the environment. Similar processes can also occur in (active) radio galaxies and galaxy clusters.

Further insight into the CR energy spectra can be gained from observations of local Galactic CRs. They are thought to be produced mainly by DSA in supernova remnants. Recently, it has become possible to observe low-energy CRs outside of the Solar System, directly in the local interstellar medium \citep[e.g.][]{Cummings2016, Stone2019}. This was done by the Voyager~1 and Voyager~2 spacecrafts when they crossed the heliopause, the place where the solar wind is stopped by the interstellar medium. The results of Voyager~2 observations of CR protons and electrons together with positrons are presented in Figure \ref{fig:energy}a. We also show examples of higher energy particles from the International Space Station experiment Alpha Magnetic Spectrometer (AMS-02) and the high altitude balloon Cosmic Ray Energetics and Mass Experiment (CREAM). All the measurements we show have been extracted from the Space Science Data Center (SSDC) and Cosmic-Ray Data Base \citep[CRDB,][]{Maurin2023} databases. 

A number of processes can affect the CR spectra as they propagate from the ISM to Earth. The solar plasma can slow down and partially remove the CRs from the interior of the solar system, causing variations in the observed flux over the solar cycle -- an effect known as solar modulation \citep[e.g.][and references therein]{Gabici2022}. This can be seen in Figure \ref{fig:energy}a in the proton and $e^- + e^+$ AMS-02 spectra as a transition towards low energies around a few GeV. However, at higher energies (above 10\,GeV) the AMS-02 and Voyager data are not affected by this effect. 

The local interstellar CR spectra can be represented by various analytical fits \citep[e.g.][]{Schlickeiser2014, Phan2018, Gabici2022}. Here we reproduce a simple broken power law function from \citet{Phan2018}:
\begin{equation}
   f(E) = C \, \frac{E^a} { \left( 1+ \frac{E}{E_b} \right) ^b }
\end{equation} 
where $E$ is the particle kinetic energy, $C$ is a constant, $E_b$ is the spectral break position, and $a$ and $b$ are parameters that cause the spectrum to change from $E^a$ below the break to $E^{a - b}$ above the break. We adopted the values fitted by \citet{Phan2018}, adjusting only the parameter $C$ to the scale of the data we used. All the parameter values are shown in Table~\ref{tab:spectra} and the model for CR protons and $e^- + e^+$ is plotted in Figure~\ref{fig:energy}.

\begin{table}[ht!]
\centering
\caption{Parameters of the fits to the local flux of CR protons and electrons together with positrons, adopted from \citet{Phan2018}.}
\label{tab:spectra}
\begin{tabular}{lcccc}
\hline 
\hline
Species      & $C\,[(\rm{m}^{2}\, \rm{s\, sr \, GeV})^{-1}]$            & $a$   & $b$ & $E_b$\,[GeV]  \\ \hline
protons      & $5.0\times 10^4$ & 0.129    & 2.829   & 0.6245 \\ \hline
$e^- + e^+$  & $9.0\times 10^2$ & $-1.236$ & 2.033   & 0.7362 \\ \hline
\end{tabular}
\end{table}
The energy distributions depend on the CR species. For protons, the spectrum is almost flat at low energies, then it peaks around 1\,GeV, and then decays as a power law with the index around $\gamma_p=-(a-b) \approx 2.7$. The change in the shape of the spectrum is thought to be mainly due to the high ionization energy loss for low-energy CR protons and the energy-dependent transport of these particles at higher energies \citep{Ruszkowski2023}. 
For example in the magnetohydrodynamic modeling of \cite{Werhahn2021a} the assumed injection energy spectral index of 2.2 for protons and electrons through energy-dependent diffusion CR transport led to the soft proton spectral index of 2.7. The slope of the measured energy spectrum of electrons and positrons in the Milky Way is different, and these particles even dominate protons at energies below $\sim50$\,MeV. The modeling of the galactic disk by \citet{Werhahn2021a} attributed this to higher Coulomb losses for protons, which become subrelativistic below 1\,GeV, while electrons remain relativistic over the whole range of energies shown in Figure \ref{fig:energy}. Including synchrotron, bremsstrahlung, and IC losses in the modeling allowed the electron spectrum to be reproduced also at higher energies. Using similar modeling, \citet{Werhahn2021a, Werhahn2021b, Werhahn2023} explained spatially resolved gamma-ray emission and radio emission in star-forming galaxies. The matched distributions of protons and primary plus secondary electrons have different positions of maxima in their energy spectra and different slopes, similar to those observed in the Milky Way.

Electron spectra in the energy range between the Voyager and AMS-02 observations can be probed by radio observations of synchrotron radiation (see the marked frequency axis at the top of Figure \ref{fig:energy}a for the radio window for the $10\,\mu$G magnetic field strength). It is in this range that the spectra change most. Measurements of galactic synchrotron diffuse emission, e.g. at 600 and 820\,MHz show a change in the synchrotron spectral index corresponding to the CR electron energy break at about 1\,GeV \citep{Tartari2008}. The observed synchrotron index reaches $\alpha\approx 1.0$, which is in good agreement with the spectral index $\gamma_e = -(a-b)\approx 3.3$ for high energies in our simple fit to the directly measured local CR electrons. 

In Figure \ref{fig:energy}b we show the intensity of the directly observed particles scaled by $E^2$ which reflects the energy density of the galactic CRs in different energy intervals. The highest energy contributions come from particles in the range from about 0.1\,GeV to 100\,GeV. The same relationship is shown in Figure \ref{fig:energy}c but without the logarithmic scale on the ordinate axis, which better represents the contributions of the different particles to the total energy budget. The dominant influence of protons is clearly visible. On the other hand, the contribution of the $e^- + e^+$ particles is small; even changes in the position of the spectral break would have little effect on the total energy of the CRs. This conclusion is consistent with the approach of \citet{Mao2012} to determine the equipartition magnetic fields in the LMC. In this work, electrons (as well as positrons) were completely neglected in the calculation of the energy of the CRs. The magnetic fields determined in this way agreed with the estimates obtained from the Faraday rotation method.

\begin{figure}[h]
\begin{center}
\includegraphics[width=8cm]{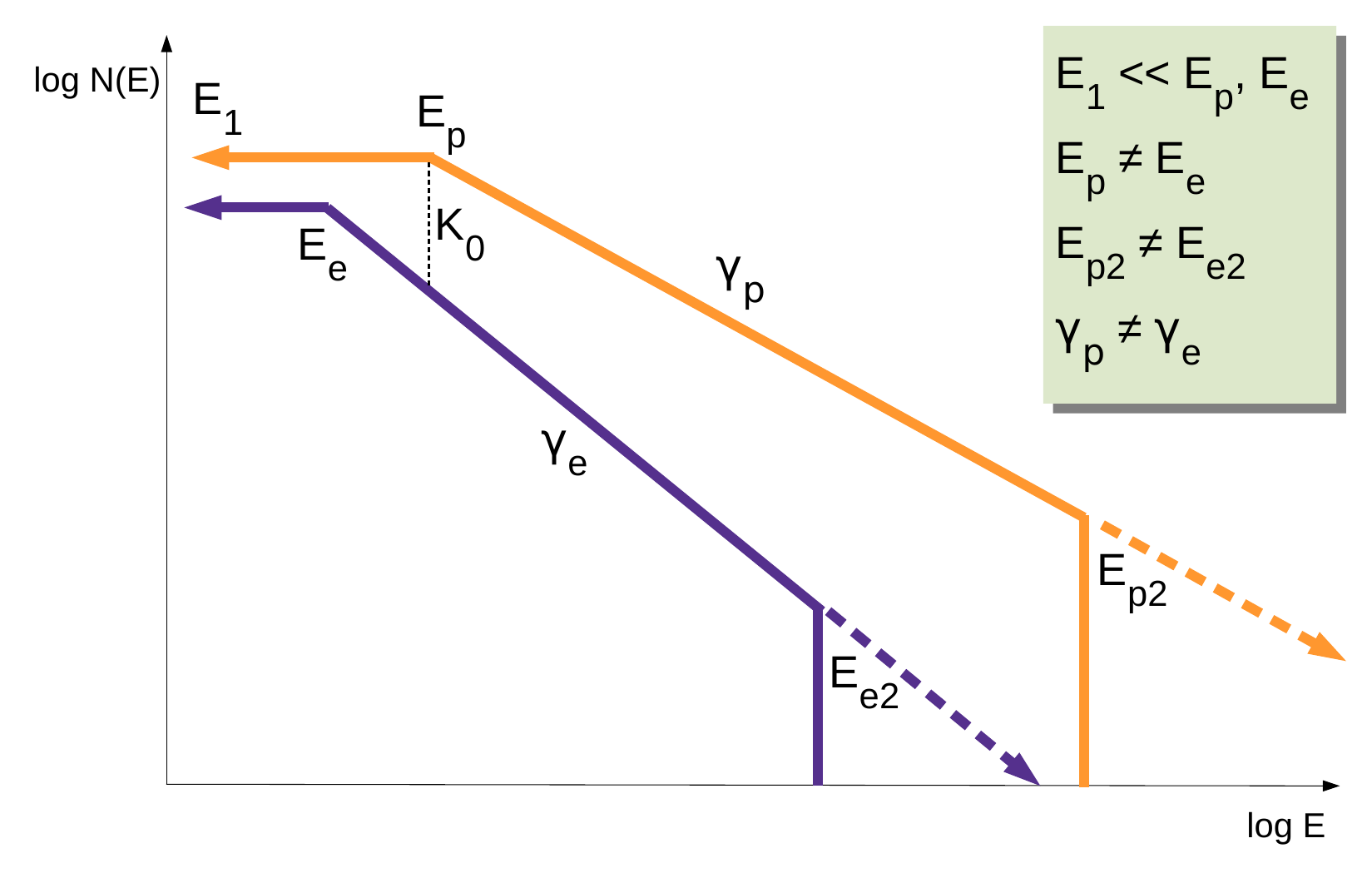}
\includegraphics[width=8cm]{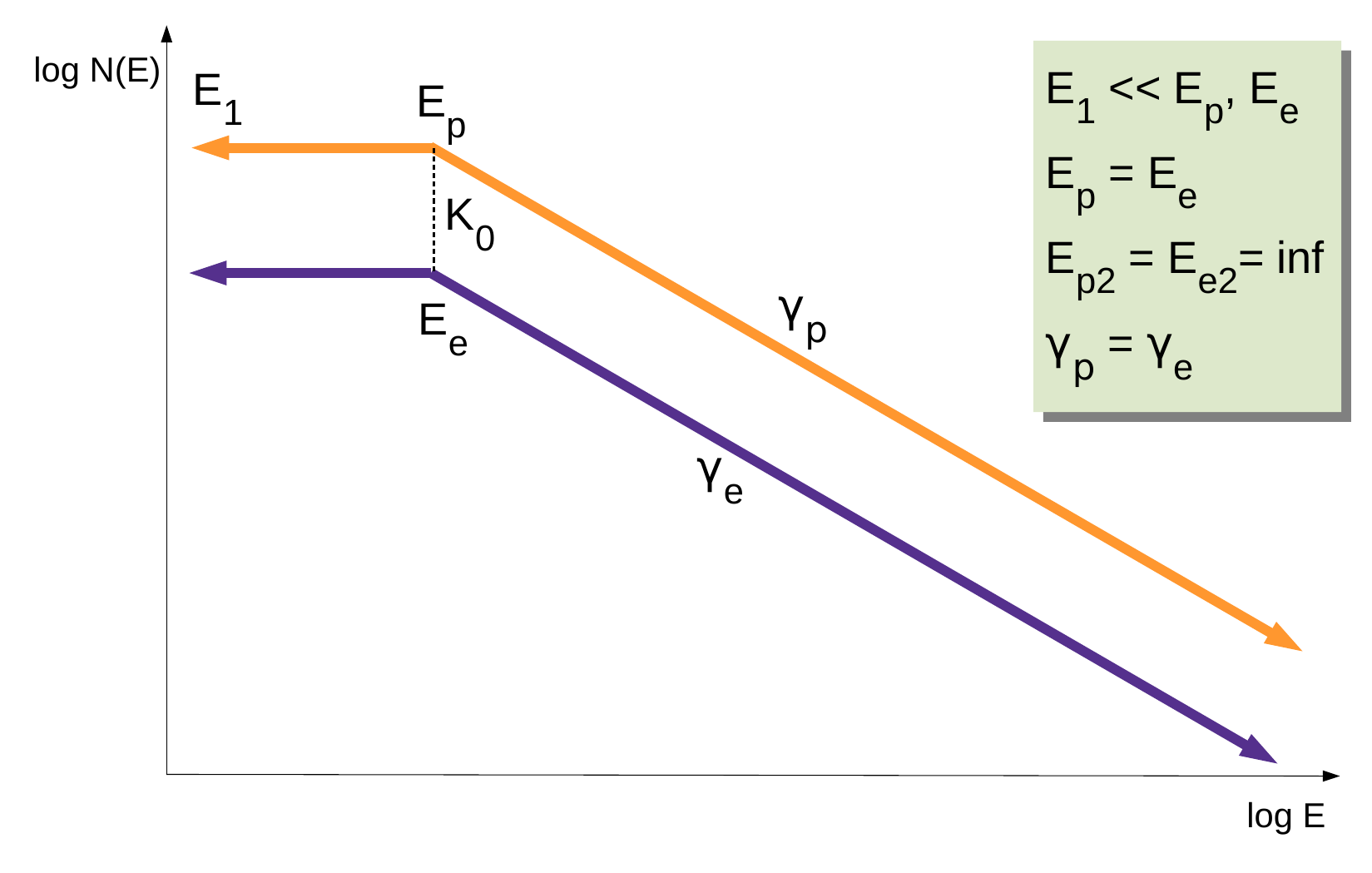}
\caption{Energy spectra used for calculation of equipartition magnetic fields: left -- possible parameter values allowed for the model used in this work; right -- frequently used CRs energy spectra for galaxies, which are a special case of the model.}
\label{fig:model_energy}
\end{center}
\end{figure}

In view of the above arguments, we propose to consider different energy spectra of the number densities of protons and electrons but in a simplified form consisting of two parts (see Figure \ref{fig:model_energy}). For protons at energies above some break energy $E_p$, the spectrum is assumed to be a power law with an index $\gamma_p$, while it remains flat in the lower part, with a low energy cutoff at some negligible energy $E_1$, and high energy cutoff at energy $E_{p2}>E_p$:
\begin{equation}
\label{eq:np}
    n_p(E)= \begin{cases}
    N_p E_p^{-\gamma_p}& E_1<E<E_p\\
    N_p E^{-\gamma_p}& E_p<E<E_{p2}\\
    \end{cases},
\end{equation}
where $N_p$ is a constant. As discussed above, the CRs spectra are not expected to be a simple power law even at high energies, and thus the model applied is only an approximate solution. The part below the energy $E_p$ could be described differently, but we leave it as a constant. This is because if we assume an approximation of the constant spectrum from e.g. $10^{-2}$\,GeV to $E_p$, and assume a rapid decrease in flux for even lower energies, then looking at Figure \ref{fig:energy}c we see that this low-energy part would bring negligible changes in the total/integrated energy of the protons. Thus, a break to constant or decay makes little difference to the energy budget once the plateau in the spectrum is sufficiently broad. 

Similarly to protons, the electron number density energy spectrum we describe as:
\begin{equation}
\label{eq:ne}
    n_e(E)= \begin{cases}
    N_e E_e^{-\gamma_e}& E_1<E<E_e\\
    N_e E^{-\gamma_e}& E_e<E<E_{e2}\\
    \end{cases},
\end{equation}
where $N_e$ is a constant and $E_e$ is the break energy in the spectrum.
This form means that if we consider electron emission from the power law part of the energy spectrum, the constant $N_e$ is the same as in Equation (\ref{eq:KS18_intensity}). In the next steps, we will use the energy equipartition constraint to eliminate this constant from the final equations. We note that other shapes of the spectra and spectra of electron-positron plasma can be introduced into the calculations (see Sections~\ref {sec:diff_cutoff} and \ref{sec:recommendations}).

To calculate the total energy density of one component of CRs, the energy spectrum must be integrated over the number density distribution of that component. For example, for protons, the total energy density can be calculated~as:
\begin{equation}
\label{eq:ep}
\begin{split}
    \epsilon_p&=\int_0^\infty n_p(E) E dE =\int_{E_1}^{E_p} N_p E_p^{-\gamma_p} E dE + \int_{E_p}^{E_{p2}}  N_p E^{-\gamma_p} E dE =\\
    &=\frac{1}{2}N_p E_p^{-\gamma_p}(E_p^2-E_1^2) +N_p\frac{E_p^{2-\gamma_p} -E_{p2}^{2-\gamma_p}}{\gamma_p-2}.
\end{split}
\end{equation}

If the energy $E_1$ is low, i.e. $E_p \gg E_1$, then $(E_p^2-E_1^2)\approx E_p^2$:
\begin{equation}
\label{eq:epsilonp}
\begin{split}
\epsilon_p&=N_p  E_p^{2-\gamma_p}\frac{1}{2}+N_p\frac{E_p^{2-\gamma_p} -E_{p2}^{2-\gamma_p}}{\gamma_p-2}=\\&=N_p \left( \frac{E_p^{2-\gamma_p}\gamma_p}{2(\gamma_p-2)}- \frac{E_{p1}^{2-\gamma_p}}{\gamma_p-2}\right).
\end{split}
\end{equation}

A similar calculation leads to an analogous expression for electron energy density:
\begin{equation}
\label{eq:ene}
\epsilon_e=N_e \left( \frac{E_e^{2-\gamma_e}\gamma_e}{2(\gamma_e-2)}- \frac{E_{e2}^{2-\gamma_e}}{\gamma_e-2}\right).
\end{equation}
In order to calculate the total energy density, we have to impose a condition on the relationship between the proton and the electron energy spectrum. Since the low-energy break in electron spectra is expected to occur at lower energies than in proton spectra due to the more efficient Coulomb cooling of protons \citep{Werhahn2023}, we assume $E_p\geq E_e$, which is also observed in the Milky Way (Figure~\ref{fig:energy}). We then define the constant $K_0$ as the ratio of the number densities of protons and electrons at energy $E_p$. That is:
\begin{equation}
    K_0=\frac{n_p(E_p)}{n_e(E_p)}=\frac{N_p}{N_e} E_p^{-(\gamma_p-\gamma_e)}.
\end{equation}
This gives the relationship between $N_p$ and $N_e$:
\begin{equation}
    \label{eq:KpAsKe}
    N_p=K_0 N_e E_p^{\gamma_p-\gamma_e}.
\end{equation}
Such a relation between the total energy densities of protons and electrons leads to the total energy density of CRs:
\begin{equation}
\label{eq:cr}
\epsilon_{cr}=\epsilon_{p}+\epsilon_{e}=K_0 N_e E_p^{\gamma_p-\gamma_e} \left( \frac{E_p^{2-\gamma_p}\gamma_p}{2(\gamma_p-2)}- \frac{E_{p2}^{2-\gamma_p}}{\gamma_p-2}\right)+N_e \left( \frac{E_e^{2-\gamma_e}\gamma_e}{2(\gamma_e-2)}- \frac{E_{e2}^{2-\gamma_e}}{\gamma_e-2}\right).
\end{equation}

\subsubsection{Equipartition formula}
\label{sec:equipartition}

Finally we use the equipartition between the CR total energy density and the energy density stored in the magnetic field, that is:
\begin{equation}
 \epsilon_{cr}=\epsilon_{B}=\frac{B_{eq}^2}{8\pi},   
\end{equation}
which together with Equation (\ref{eq:cr}) gives:
\begin{equation}
\epsilon_{p}+\epsilon_{e}=K_0 N_e E_p^{\gamma_p-\gamma_e} \left( \frac{E_p^{2-\gamma_p}\gamma_p}{2(\gamma_p-2)}- \frac{E_{p2}^{2-\gamma_p}}{\gamma_p-2}\right)+N_e \left( \frac{E_e^{2-\gamma_e}\gamma_e}{2(\gamma_e-2)}- \frac{E_{e2}^{2-\gamma_e}}{\gamma_e-2}\right)=\frac{B_{eq}^2}{8 \pi}.
\end{equation}
Since $B_{eq}$ denotes the average magnetic field within a significant volume containing numerous orientations of $B_r$, it is also satisfied that $B_{eq}^2=B_u^2+B_r^2$. Using Equation (\ref{eq:KpAsKe}) and solving for the constant $N_e$ we obtain:
\begin{equation}
\label{eq:N_e}
N_e=\left[K_0   \frac{E_p^{2-\gamma_e}\gamma_p}{2(\gamma_p-2)}\left(1- \frac{2}{\gamma_p}\left(\frac{E_{p2}}{E_p}\right)^{2-\gamma_p}\right)+ \frac{E_e^{2-\gamma_e}\gamma_e}{2(\gamma_e-2)} \left(1-\frac{2}{\gamma_e}\left(\frac{E_{e2}}{E_e} \right)^{2-\gamma_e}\right) \right]^{-1} \frac{B_{eq}^2}{8 \pi}
\end{equation}
Only the last term of this formula depends on the magnetic field. Finally, we use this expression to replace the constant $N_e$ in Equations (\ref{eq:KS18_intensity}) to obtain:
\begin{equation}
\label{eq:Ieq}
    I_\nu= \frac
      {
      \frac{B_{eq}^2}{8 \pi} \left(\frac{\nu}{2c_1}\right)^\frac{1-\gamma_e}{2} \frac{c_2 f l}{4 \pi} \int_{4\pi}(B\sin{\mu})^{\frac{\gamma_e+1}{2}}d\Omega
      }
      {K_0 \frac{E_p^{2-\gamma_e}\gamma_p}{2(\gamma_p-2)}\left(1- \frac{2}{\gamma_p}\left(\frac{E_{p2}}{E_p}\right)^{2-\gamma_p}\right)+ \frac{E_e^{2-\gamma_e}\gamma_e}{2(\gamma_e-2)} \left(1-\frac{2}{\gamma_e}\left(\frac{E_{e2}}{E_e} \right)^{2-\gamma_e}\right)
      }
\end{equation}

\begin{equation}
\label{eq:Peq} 
    {PI}_\nu=\frac
    {
    \frac{B_{eq}^2}{8 \pi} \left(\frac{\nu}{2c_1}\right)^\frac{1-\gamma_e}{2} \frac{p_0 c_2 f l}{4 \pi} \int_{4\pi}(B\sin{\mu})^{\frac{\gamma_e+1}{2}} \cos{2\chi}d\Omega
    }
    {
    K_0  \frac{E_p^{2-\gamma_e}\gamma_p}{2(\gamma_p-2)}\left(1- \frac{2}{\gamma_p}\left(\frac{E_{p2}}{E_p}\right)^{2-\gamma_p}\right)+ \frac{E_e^{2-\gamma_e}\gamma_e}{2(\gamma_e-2)} \left(1-\frac{2}{\gamma_e}\left(\frac{E_{e2}}{E_e} \right)^{2-\gamma_e}\right) 
    }.
\end{equation}

The derived formulas given by Equations (\ref{eq:Ieq}) and (\ref{eq:Peq}) are used to calculate synchrotron intensities under the assumption of energy equipartition in our Bayesian approach.

\subsection{Special cases}
\label{sec:special_cases}

In general, formulas in Equations (\ref{eq:Ieq}) and (\ref{eq:Peq}) cannot be inverted to recover either the total magnetic field strength or its components. However, in special cases of a purely random or purely uniform magnetic field, these formulas reduce to invertible forms.

\subsubsection{Purely uniform field}

In the presence of only a purely uniform magnetic field component, we have $B_{eq}=B_{u}$, and $B\sin{\mu}=B_{u} \sin{\theta_0}$. In this case, the integrand of Equation (\ref{eq:Ieq}) does not depend on the orientation of the random field component giving:
\begin{equation}
     \frac{1}{4 \pi}\int_{4\pi} (B \sin{\mu})^{\frac{\gamma_e+1}{2}}d\Omega= \frac{1}{4 \pi}\int_{4\pi} (B_{u} \sin{\theta_0})^{\frac{\gamma_e+1}{2}}d\Omega= (B_{u} \sin{\theta_0})^{\frac{\gamma_e+1}{2}}.
\end{equation}
Substituting this into Equation (\ref{eq:Ieq}) we get:
\begin{equation}
\label{eq:Ieq_Bu}
    I_\nu= \frac
      {
      \frac{B_{u}^2}{8 \pi} \left(\frac{\nu}{2c_1}\right)^\frac{1-\gamma_e}{2} c_2 f l \left(B_{u} \sin{\theta_0}\right)^{\frac{\gamma_e+1}{2}}
      }
      {K_0   \frac{E_p^{2-\gamma_e}\gamma_p}{2(\gamma_p-2)}\left(1- \frac{2}{\gamma_p}\left(\frac{E_{p2}}{E_p}\right)^{2-\gamma_p}\right)+ \frac{E_e^{2-\gamma_e}\gamma_e}{2(\gamma_e-2)} \left(1-\frac{2}{\gamma_e}\left(\frac{E_{e2}}{E_e} \right)^{2-\gamma_e}\right)
      },
\end{equation}
which can be solved for $B_{u}$:
\begin{equation}
\label{eq:Bu_only}
B_{u}=\biggl\{\frac{8\pi \left(\frac{\nu}{2c_1}\right)^\frac{\gamma_e-1}{2}I_\nu}{\left(\sin{\theta_0}\right)^{\frac{\gamma_e+1}{2}} c_2 f l} \left[K_0  \frac{E_p^{2-\gamma_e}\gamma_p}{2(\gamma_p-2)}\left(1- \frac{2}{\gamma_p}\left(\frac{E_{p2}}{E_p}\right)^{2-\gamma_p}\right)+ \frac{E_e^{2-\gamma_e}\gamma_e}{2(\gamma_e-2)} \left(1-\frac{2}{\gamma_e}\left(\frac{E_{e2}}{E_e} \right)^{2-\gamma_e}\right)\right]\biggr\}^{\frac{2}{\gamma_e+5}}.
\end{equation}

For $B_r=0$ from Equation (\ref{eq:sin}) we get $\cos{2\chi}=1$, which from Equation (\ref{eq:Peq}) leads to polarized intensity $PI_\nu=p_0 I_\nu$.

\subsubsection{Purely random field}
\label{sec:purely_random}
In the case of a purely random magnetic field, we have in Equation (\ref{eq:Ieq}) $B_{eq}=B_{r}$, and $B\sin{\mu}=B_{r} \sin{\theta}$. The integral over the solid angle has an analytic solution:
\begin{align}
\label{eq:sin_avg}
\begin{split}
    \frac{1}{4\pi}\int_{4\pi}(B_{r}\sin{\theta})^{\frac{\gamma_e+1}{2}}d\Omega&= \frac{1}{4\pi}\int_{0}^{2 \pi} \int_{0}^{\pi} (B_{r} \sin{\theta})^{\frac{\gamma_e+1}{2}} \sin{\theta} d\theta d\phi = \\ &=\frac{1}{2}B_{r}^{\frac{\gamma_e+1}{2}}\; \int_{0}^{\pi}(\sin{\theta})^{\frac{3+\gamma_e}{2}} d\theta=\\&=  B_{r}^{\frac{\gamma_e+1}{2}} \frac{\sqrt{\pi}}{2} \frac{\Gamma\left(\frac{\gamma_e+5}{4}\right)}{\Gamma\left(\frac{\gamma_e+7}{4}\right)}.
\end{split}
\end{align}

An analogous formula for averaging field directions is given, for example, in \citet{tools}. Substituting it into Equation (\ref{eq:Ieq}) gives intensity in the case of purely random magnetic field:

\begin{equation}
\label{eq:Ieq_Br}
    I_\nu= \frac
      {
      \frac{B_{r}^2}{8 \pi} \left(\frac{\nu}{2c_1}\right)^\frac{1-\gamma_e}{2} c_2 f l  B_{r}^{\frac{\gamma_e+1}{2}} \frac{\sqrt{\pi}}{2} \frac{\Gamma\left(\frac{\gamma_e+5}{4}\right)}{\Gamma\left(\frac{\gamma_e+7}{4}\right)}
      }
      {K_0 \frac{E_p^{2-\gamma_e}\gamma_p}{2(\gamma_p-2)}\left(1- \frac{2}{\gamma_p}\left(\frac{E_{p2}}{E_p}\right)^{2-\gamma_p}\right)+ \frac{E_e^{2-\gamma_e}\gamma_e}{2(\gamma_e-2)} \left(1-\frac{2}{\gamma_e}\left(\frac{E_{e2}}{E_e} \right)^{2-\gamma_e}\right)
      },
\end{equation}
Inverting this formula to obtain the strength of magnetic field just as we have done for the uniform field, we get: 
\begin{equation}
\label{eq:Br_only}
B_{r}=\biggl\{\frac{8\pi \left(\frac{\nu}{2c_1}\right)^\frac{\gamma_e-1}{2} I_\nu} { c_2 f l} 
\left[K_0 \frac{E_p^{2-\gamma_e} \gamma_p   } {2 \left( \gamma_p - 2 \right) } \left(1- \frac{2}{\gamma_p}\left(\frac{E_p}{E_{p2}}\right)^{\gamma_p -2}\right) + 
\frac{E_e^{2-\gamma_e} \gamma_e }{2 \left( \gamma_e-2 \right) } \left(1-\frac{2}{\gamma_e}\left(\frac{E_e}{E_{e2}} \right)^{\gamma_e -2}\right)\right] \frac{2}{\sqrt{\pi}} \frac{\Gamma\left(\frac{\gamma_e+7}{4}\right)}{\Gamma\left(\frac{\gamma_e+5}{4}\right)} \biggr\}^{\frac{2}{\gamma_e+5}}.
\end{equation}

In the case of a purely random field the integrand of Equation (\ref{eq:Peq}) vanishes, giving $PI_\nu=0$.

\section{Bayesian approach to equipartition}
\label{sec:bayesian}
Bayesian methods involve postulating the prior probability densities and determining a likelihood function from which the (joint) posterior distribution of the modeled parameters can be computed. MCMC methods are widely used in Bayesian inference and provide powerful algorithms for sampling the modeled parameter space. In this section, we will discuss the set of parameters we have chosen for our implemented Bayesian model and construct the posterior. We treat the observed values of intensity ${I_\nu}_{obs}$ and ${PI_\nu}_{obs}$ as data $\bm{D}$, and for $\bm{\theta}$ we include four parameters in our model: the spectral index $\alpha$ of the synchrotron radiation, $K_0$, and the desired magnitudes of two magnetic field components $B_u$ and $B_r$, which give the total magnetic field strength $B_t=\sqrt{B_u^2+B_r^2}$. Our model, which allows us to calculate the intensities $I_\nu(B_u,B_r,\alpha,l,K_0)$ and $PI_\nu(B_u,B_r,\alpha,l,K_0)$ is given by Equations (\ref{eq:Ieq}) and (\ref{eq:Peq}). Note that although $\alpha$ is also a quantity measured by observations, it is treated as a model parameter, and its measurements are used as a source of a prior probability distribution. In this particular Bayesian model, all other quantities appearing in Formulas (\ref{eq:Ieq}) and (\ref{eq:Peq}), such as frequency $\nu$ and path length $l$ are part of the model $\bm{M}$, take fixed values and are assumed here to be known.

\subsection{Prior distributions}
\label{sec:prior}

To proceed with Bayesian inference, we need to start with some prior probability distributions for the parameters. These are assumed distributions describing our knowledge of the values of parameters before taking into account intensity observations ${I_\nu}_{obs}$ and ${PI_\nu}_{obs}$. For magnetic field components $B_u$ and $B_r$ we choose simple uninformative priors in the form of uniform distributions $U(B_u,B_{min},B_{Max})$ and $U(B_r,B_{min},B_{Max})$ from $B_{min}=0$ $\mu$G to some sufficiently large upper bound, e.g. $B_{max}=1000$ $\mu$G. An alternative Jeffreys prior is discussed in Appendix \ref{sec:app_prior}. The spectral index prior is modeled as a truncated Gaussian $N_T(\alpha,\alpha_{obs},\sigma_\alpha^2)$ where the mean value $\alpha_{obs}$ and the standard deviations $\sigma_\alpha$ are the measured observed spectral index and its observational uncertainty. The truncation is to limit the values of $\alpha$ above $\alpha_{min}$ equal to 0.0 or 0.5, to ensure the validity of the derived formulas (see Appendix \ref{sec:cutoff}). For the path length $l$ we propose a weakly informative prior in the form of a truncated Gaussian prior $N_t(l,\mu_l,\sigma^2_l)$, restricted to $l>0$. Similarly, for $K_0$ we include a truncated Gaussian prior $N_t(K_0,\mu_{K_0},\sigma_{K_0}^2)$, restricted to values of $K_0\geq0$. The expected value of $K_0$ for normal galaxies can be taken as $\mu_{K_0}=100$ (Section~\ref{sec:introduction}), with some standard deviation, e.g. $\sigma_{K_0}=10$, representing our lack of precise knowledge of $K_0$. Hence, the joint prior distribution derived as a product of priors on the individual parameters takes the form:
\begin{equation}
\pi_{\text{prior}}(B_u,B_r,\alpha, l,K_0)=
    C
    \frac{1}{\sqrt{2\pi}\sigma_\alpha} \exp\left[-\frac{\left(\alpha-\alpha_{obs}\right)^2}{2\sigma_\alpha^2} \right]
    \frac{1}{\sqrt{2\pi}\sigma_{l}} \exp\left[-\frac{\left(l-\mu_l\right)^2}{2\sigma_{l}^2} \right]
    \frac{1}{\sqrt{2\pi}\sigma_{K_0}} \exp\left[-\frac{\left(K_0-\mu_{K_0}\right)^2}{2\sigma_{K_0}^2} \right],
\end{equation}
which remains valid for $B_u\in (B_{min},B_{max})$, $B_r\in (B_{min},B_{max})$, $\alpha\in (\alpha_{min},\infty)$, $l \in (0,\infty)$, $K_0 \in (0,\infty)$, and becomes zero everywhere else. $C$ is the normalization constant. The exact value of $C$ is not required to proceed with the MCMC methods.

\subsection{Likelihood}

The likelihood function is constructed from a pair of measurements: one of the total synchrotron intensity ${I_\nu}_{obs}$ and one of the polarized intensity ${PI_\nu}_{obs}$. If both are modeled as noisy measurements with observation uncertainties described by the Gaussian noise distribution, then the likelihood of measuring the values of $\bm{D}=({I_\nu}_{obs},{PI_\nu}_{obs})$ given values of parameters $\bm{\theta}=(B_u,B_r,\alpha,l,K_0)$ is:
\begin{equation}
\begin{split}
    & \pi_{\text{likelihood}}({I_\nu}_{obs},{PI_\nu}_{obs}|B_u,B_r,\alpha,l,K_0)=\\
    & \frac{1}{\sqrt{2\pi}\sigma_I} \exp\left[-\frac{\left({I_\nu}_{obs}-I_\nu(B_u,B_r,\alpha,l,K_0)\right)^2}{2\sigma_I^2} \right] \frac{1}{\sqrt{2\pi}\sigma_{PI}} \exp\left[-\frac{\left({PI_\nu}_{obs}-{PI}_\nu(B_u,B_r,\alpha,l,K_0)\right)^2}{2\sigma_{PI}^2} \right]
\end{split}
\end{equation}

\subsection{Posterior distribution}
After assuming the prior distribution, and determining the likelihood function, the posterior distribution according to Equation (\ref{eq:bayesian_general}) is:
\begin{equation}
\label{eq:posterior}
    \pi_{\text{posterior}}(B_u,B_r,\alpha,l,K_0 | {I_\nu}_{obs},{PI_\nu}_{obs}) = \frac{\pi_{\text{likelihood}}({I_\nu}_{obs},{PI_\nu}_{obs} | B_u,B_r,\alpha,l,K_0) \, \pi_{\text{prior}}(B_u,B_r,\alpha,l,K_0)} { \pi_{\text{evidence}}( {I_\nu}_{obs},{PI_\nu}_{obs})},
\end{equation}
where $\pi_{\text{evidence}}( {I_\nu}_{obs},{PI_\nu}_{obs})$ plays the role of normalization constant for the distribution. It is not necessary (and usually very hard) to determine it. MCMC methods used to generate samples require only a function proportional to the probability density function to be known. This means that only the numerator of Equation (\ref{eq:posterior}) is needed to proceed with the calculations.

\section{Example of computation of a posteriori distribution of magnetic field strength} 
\label{sec:example}

In this section, we demonstrate an example application of the Bayesian model introduced in the previous section. To do this, we have chosen a set of parameters representing a fiducial region observed in a nearby galaxy at a high frequency of 4.86\,GHz. The selected values of the physical parameters and observables are given in Table \ref{tab:example_input}. Such values of synchrotron total and polarized emission at $15\arcsec$ resolution can be found, for example, north and south of the pseudoring in the galaxy NGC\,4736 \citep{ChyzyButa2008} or in the southern polarized ridge in NGC\,4254 \citep{Chyzy2008_NGC4254SFR}. For the synchrotron path length we applied a typical value of 1\,kpc \cite[e.g.][]{Fletcher2011, Drzazga2011} with a 10\% uncertainty. We assume that the inclination $i$ of the uniform field with respect to the plane of the sky, $i=90\degr-\theta_0$, is $45\degr$ in the fiducial region.

\begin{table}[h]
\centering
\caption{Values of physical parameters and observables for the high-frequency fiducial region.}
\label{tab:example_input}
\begin{tabular}{lcc}
\hline
\hline
Parameter& Unit & Value  \\ \hline
$I_{obs}$&mJy/beam                     &   1.0       \\ \hline
$\sigma_{I_{obs}}$&mJy/beam              &    0.1       \\ \hline
$PI_{obs}$&mJy/beam                               & 0.2          \\ \hline        
$\sigma_{QU}$&mJy/beam                                        & 0.02          \\ \hline
$\alpha_{obs}$ & -                      &  1.0         \\ \hline
$\sigma_{\alpha}$ & -                      &  0.1         \\ \hline
$\nu$  & GHz                      & 4.86          \\ \hline
$\theta_{min}$& arcsec & 15 \\ \hline
$\theta_{maj}$& arcsec & 15 \\ \hline
$l$& pc & 1000 \\ \hline
$\sigma_{l}$ & pc& 100 \\ \hline
$K_0$& -& 100 \\ \hline
$\sigma_{K_0}$& -& 10 \\ \hline
$i$& deg & 45 \\ \hline
$E_p=E_e$ & GeV & 0.938 \\ \hline
$E_{p2}=E_{e2}$ & GeV & $\infty$ \\ \hline
$f$ & - & 1 \\ \hline
\end{tabular}
\end{table}

We write a generic Monte Carlo sampler to explore the parameter space that uses the Metropolis-Hastings (M-H) algorithm to generate samples from the posterior distribution described by Equation (\ref{eq:posterior}). We start by initializing the chains. The joint maximum of the multivariate posterior probability (MAP) estimator was used to find approximate values of the search parameters. We then spread these values over a small Gaussian sphere to find the starting points of the MCMC chains. This procedure allows us to start chains in regions of non-negligible probability density, allowing for faster convergence compared to starting far away from the maximum of the multivariate posterior distribution.

\begin{figure}[h]
\begin{center}
\includegraphics[width=8.9cm]{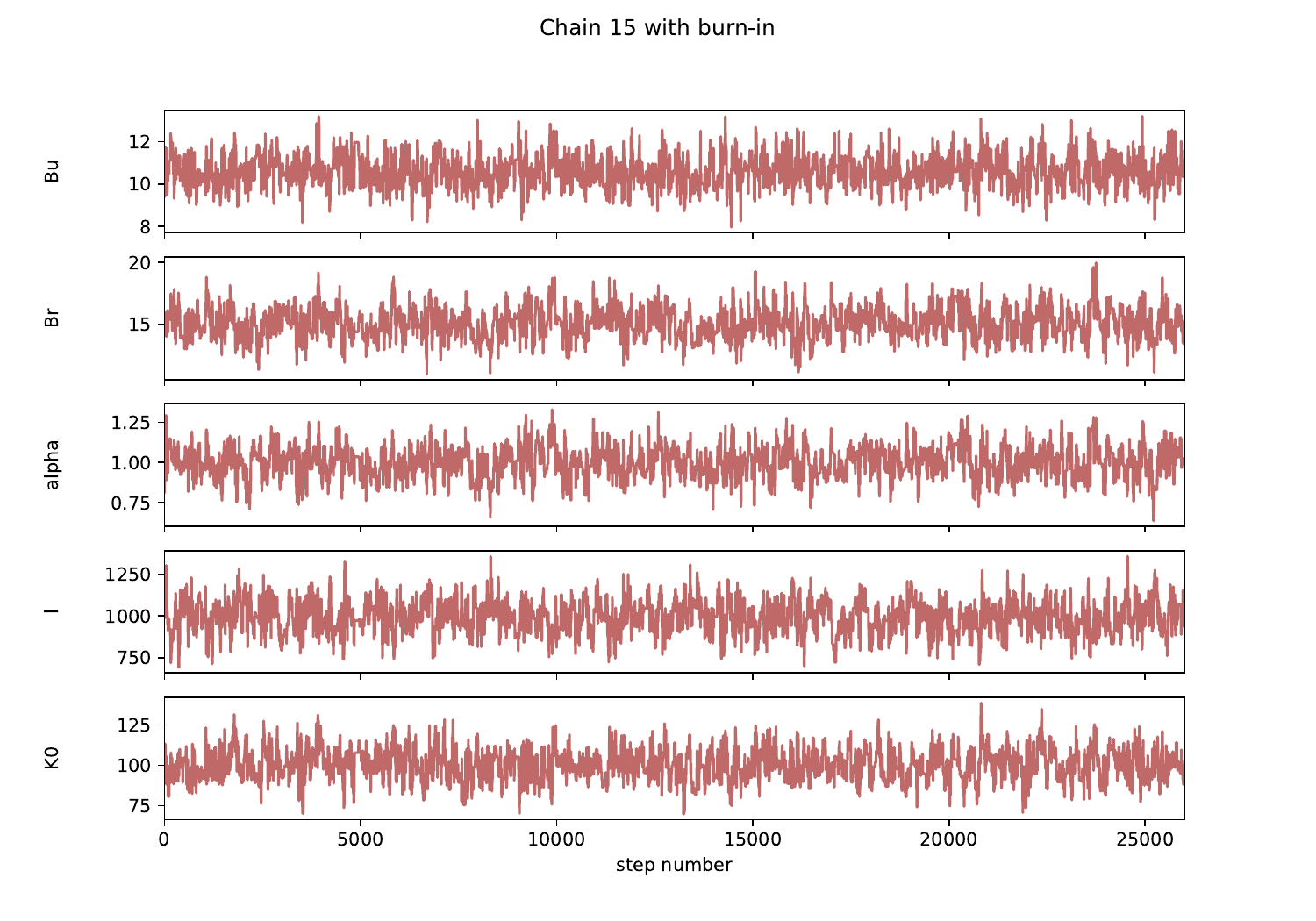}
\includegraphics[width=8.9cm]{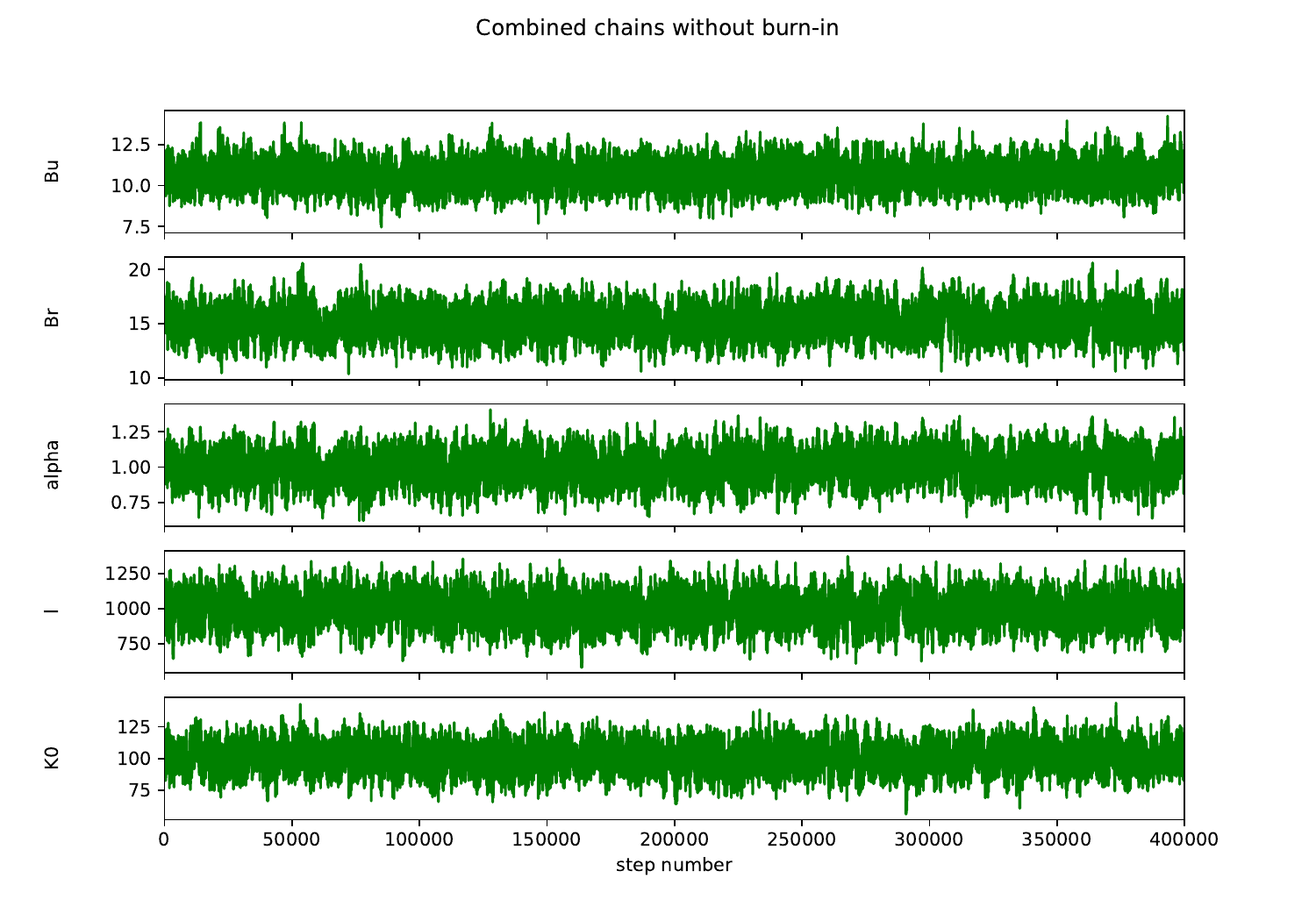}
\caption{Left: An example of a single  trace plot of an individual chain (walker) for the parameters from the M-H method. Right: A trace plot of the modeled parameters for 16 merged chains and removed burn-in steps.}
\label{fig:mh_traceplot}
\end{center}
\end{figure}

\begin{figure}[h]
\epsscale{1.3}
\includegraphics[width=18cm]{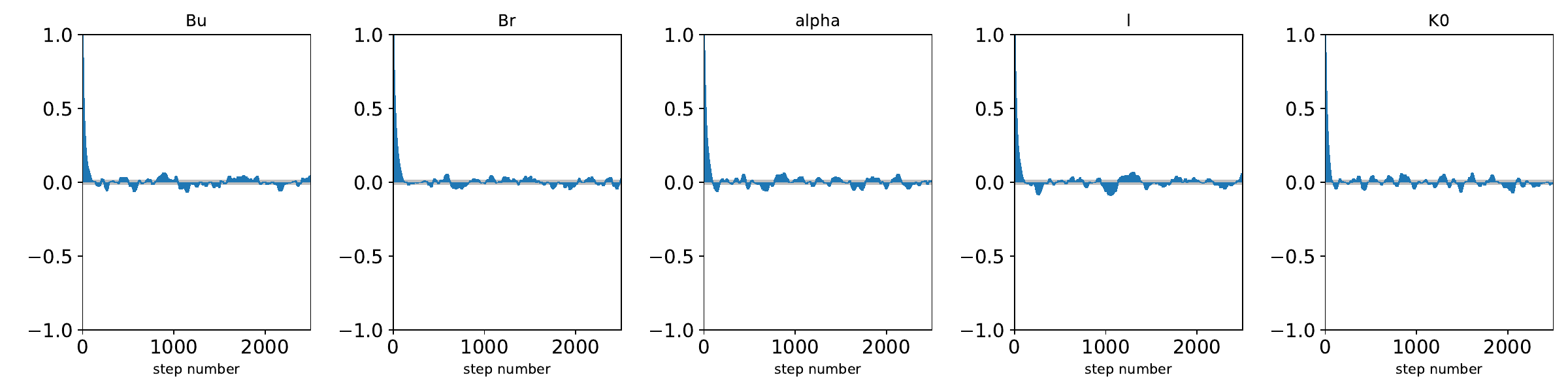}
\caption{Autocorrelation plots for parameters in a single chain from the MCMC M-H method.}
\label{fig:mh_correl}
\end{figure}

We run the code with 16 chains of 26,000 steps each and a burn-in of 1,000 steps per chain. Figure \ref{fig:mh_traceplot} shows a trace plot of a single walker as a function of the step in the chain and a similar trace plot for all merged chains with the burn-in parts removed. In Figure \ref{fig:mh_correl} we present an autocorrelation plot for a single chain. It is clear from these figures the walkers begin to explore the posterior distribution and quickly (after less than 500 steps) stabilize the scattering. The samples are randomly distributed around the mean with a constant bound and keep the autocorrelation remains low. The rank-normalized $\hat{R}$ statistics was used as an additional diagnostic test for the convergence of walkers. Values below the recommended threshold of 1.01 \citep{Vehtari2021} indicate that the chains have converged, walkers have sufficiently explored the space and the sample generated has reached the ultimate distribution. For the sample presented in this example, the $\hat{R}$ values were 1.0015, 1.0042, 1.0051, and 1.0014, 1.0019 for $B_u$, $B_r$, $\alpha$, $l$, $K_0$ respectively, indicating that the MCMC process easily and quickly converged to the target distribution and provided a good representation of the posterior.

\begin{figure}[h]
\includegraphics[width=0.9\textwidth]{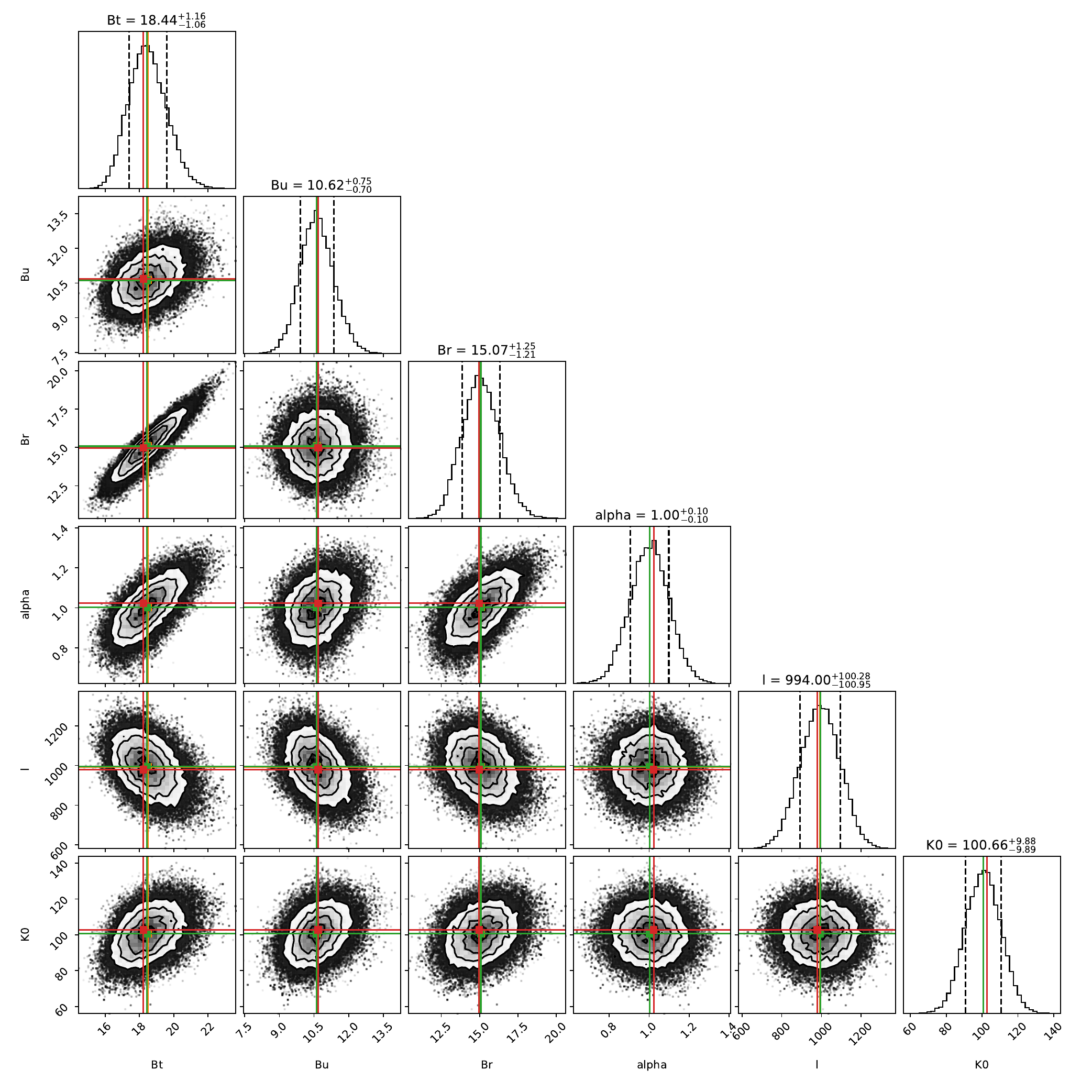}
\caption{Corner plot of MCMC samples from the M-H method showing posterior distributions of magnetic field components and model parameters with with mean (orange), median (green), and mode (red) vertical lines. Dashed vertical lines represent 68\% credible intervals. The contours of the 2D marginalized posterior distributions contain 11.8\%, 39.3\%, 67.5\% and 86.4\% of the samples corresponding to 1, 1.5, 2, 2.5 standard deviation  equivalents. The median values of the parameters and the uncertainties given by the 68\% credible intervals are shown at the top of the histograms.}
\label{fig:mh_corner}
\end{figure}

In the Bayesian paradigm, uncertainties in each model parameter are naturally propagated to uncertainties of derived parameters, as in the formula for the total magnetic field strength $B_t=\sqrt{B_u^2 + B_r^2}$. Therefore, we include this parameter in the following statistical analysis of the magnetic field strength. The corner plot presented in Figure \ref{fig:mh_corner} shows all one- and two-dimensional projections of the posterior probability distributions of our main Bayesian parameters $B_u$, $B_r$, the auxiliary parameters $\alpha$, $l$, $K_0$, and the transformed parameter $B_t$. This plot visualizes all the covariances between parameters and the marginalized distributions of each parameter as histograms. To characterize the parameters obtained from marginalization of the posterior, for all of them, we calculated three different point estimates and credible intervals. They are defined as follows:

\begin{itemize}
    \item the mean estimated as the average of the sampled points, and centered credible interval (CCI) with equal probability of 34\% above and below the mean value;
    \item the median and the symmetric credible interval (SCI) spanning from the 16th to the 84th percentile;
    \item the mode estimated using the half-sample mode algorithm \citep{Half_Sample_mode} and the highest density interval (HDI), i.e., the minimum width interval containing 68\% of the probability.
\end{itemize}
The obtained estimates are presented in Table \ref{tab:mh_estimates}. All derived intervals contain 68\% of the samples, estimating the $1\sigma$ equivalent of uncertainty. We see that none of these intervals is necessarily symmetrical about its respective point estimate. The resulting values for the mean, median, and mode are very similar, differing by much less than 10\% of their uncertainties.

\begin{table}[h!]
\caption{Point estimates and 68\% credible intervals of the marginal posterior probability distributions for the Bayesian parameters using the MCMC M-H method.}
\label{tab:mh_estimates}
\begin{tabular}{lcrcrcrc}
\hline
\hline
Parameter& Unit & mean & mean CCI & median & median SCI & mode  & mode HDI\\ \hline
$B_t$ & $\mu$G & 18.49& 17.46 -- 19.70& 18.44& 17.39 -- 19.60& 18.23& 17.38 -- 19.60\\ \hline 
$B_u$ & $\mu$G & 10.64&  9.95 -- 11.41& 10.62&  9.91 -- 11.36& 10.67&  9.85 -- 11.30\\ \hline 
$B_r$ & $\mu$G & 15.10& 13.90 -- 16.37& 15.07& 13.86 -- 16.32& 14.96& 13.85 -- 16.31\\ \hline 
$\alpha$ & &  1.00&  0.90 --  1.10&  1.00&  0.90 --  1.10&  1.02&  0.91 --  1.10\\ \hline 
$l$&pc&993&892 -- 1093&994&893 -- 1094&979&890 -- 1091\\ \hline 
$K_0$ & &100.7& 90.8 -- 110.5 &100.7 & 90.87 -- 110.6 & 102.7 & 90.7 -- 110.5\\ \hline 
\end{tabular}
\end{table}

Although the parameters $\alpha$, $l$, and $K_0$ are shown in the corner plot in Figure \ref{fig:mh_corner} and in Table \ref{tab:mh_estimates}, the parameters in which we are mainly interested in are $B_u$, $B_r$, and the transformed parameter $B_t$, so the marginal distributions related to these parameters are of interest to us. We note that the marginal distributions of $\alpha$, $l$, and $K_0$ closely resemble their informative Gaussian priors and appear to be uncorrelated with each other, which is to be expected since there is no reason to use the measurements of synchrotron intensity to significantly improve our knowledge of either $\alpha$, $l$ or $K_0$. The procedure of marginalizing out parameters that are required in Bayesian inference because they are real sources of uncertainty but are not of the primary interest is explained, for example, in \citet{tutorial_on_Bayesian}. The two-dimensional marginal distributions of $\alpha$ paired with all magnetic field strength parameters show positive correlations. Stronger magnetic fields are more likely for higher spectral indexes in the range allowed by the chosen prior on $\alpha$. The impact of the spectral index on the determined magnetic field strength is discussed in Section \ref{sec:diff-alpha}. Similarly, $B_t$ is positively correlated with both of its components $B_u$ and $B_r$. The correlation of $B_t$ with $B_r$ is visually much more apparent than the correlation with $B_u$ because the recovered marginalized distribution of $B_r$ spans larger range of magnetic field strength than that of $B_u$. The spread of the distributions of the magnetic field strength parameters is a direct result of propagation of uncertainties in the observed intensities and other parameters. These effects are discussed in more detail in Section \ref{sec:comp}.

\section{Discussion}
\label{sec:discussion}

\subsection{Different methods to generate MCMC}
\label{sec:EMCEE}

In this study, we used a Bayesian method to determine magnetic field strengths based on the Metropolis-Hastings MCMC algorithm for sampling target distributions. How reliable, accurate, and stable are the results obtained in this way? To have independent results for comparison, we developed an alternative code using an algorithm based on the open-source PYTHON package EMCEE \citep{EMCEE_python}, an implementation of the Affine-Invariant (A-I) MCMC Ensemble sampler \citep{EMCEE_algorytm}. We used exactly the same values of the observational parameters given in Table \ref{tab:example_input} to set priors and start the MCMC simulations. We found that EMCEE required several times shorter chains than our Metropolis-Hasting code to converge, but at the cost of several times longer run time. We present results for running 16 walkers for 16,000 steps each, giving a total of 240,000 steps without burn-in. We have checked and verified the correct convergence of the chains using the rank-normalized $\hat{R}$ diagnostic test. 

The resulting posterior distributions of the magnetic fields are shown in Figure~\ref{fig:em_corner} (Appendix \ref{sec:app_figures}) which should be compared with Figure~\ref{fig:mh_corner} from our M-H code. Figures \ref{fig:em_traceplot}, \ref{fig:em_correl}, present the chain trace plots and the autocorrelation plot illustrating the correct convergence of the walkers to the stationary target distributions. The two codes are in complete agreement. In particular, for the median estimates of $B_t$, $B_u$, $B_r$ we obtained for the A-I code $18.48^{+1.18}_{-1.09}$, $10.59^{+0.72}_{-0.70}$, and $15.13^{+1.29}_{-1.23}$, which differ from the values of the M-H code (Table~\ref{tab:mh_estimates}) by only 0.2\%, 0.3\%, and 0.4\%, respectively. In turn, the limits of the median SCI range differ by a maximum of 4\% (in the range of values of $B_u$. The situation is similar for the estimated mean magnetic field values. The differences are largest between the mode values, but reach only 0.8\%.  So for all point estimates of all magnetic field components, the differences are less than one order of magnitude of the estimated 68\% uncertainties of the field strength estimates. Thus, the magnetic field values obtained are independent of the posterior sampling method used.

We also compared the results from the M-H simulation with those from the same algorithm but starting with a completely different set of seed numbers to initialize the pseudo-number generators. In this case, for example, the median values of the $B_u$ and $B_r$ distributions deviated from the original values by up to 3\% demonstrating the stability of the results. These calculations indicate that, for simulation chain lengths used, the differences in the results from the M-H and A-I methods are similar to those resulting from the natural randomness implicit in the Monte Carlo methods. Of course, with both methods it is always possible to increase the number of samples generated to reach any level of convergence and further reduce fluctuations in the distribution of parameters. However, given the magnitude of the estimated uncertainties, smaller fluctuations in the point estimates become irrelevant, and for that reason spending more computational time on adding more samples leads to diminishing returns.

\subsection{Comparison of simulations with increasing uncertainties of the observations}
\label{sec:comp}

Applying the Bayesian method to the determination of equipartition magnetic fields naturally allows the determination of the uncertainty of the estimated field. We now want to investigate exactly how the different parameters and their uncertainties affect the posterior distribution of the magnetic field and the derived its positional parameters. We have modeled the local magnetic field in a single region of a galaxy observed at low frequency. Our fiducial region was inspired by a single inter-arm region in the northwestern part of the galaxy NGC\,6946 observed with LOFAR interferometer at $\nu=0.138$\,GHz \citep{Shimwell2022}. The nonthermal emission in this region of $I_{obs}=10\pm0.5$\,mJy/beam results from a slightly asymmetric beam area of $20.9\arcsec \times 20.34\arcsec$. A nonthermal spectral index $\alpha=0.76 \pm 0.02$ was determined from LOFAR map and 1.465\,GHz map \citep[from][]{Beck2007}. We assumed an effective path length through the galactic disk $l=1190\pm 50$\,pc, which is 1\,kpc disk thickness corrected for galactic inclination ($33^{\circ}$) and takes into account $3^{\circ}$ inclination uncertainty. We presumed $K_0=100\pm 10$ and set $f=1$. We then varied this fiducial model by sequentially changing the uncertainty values of the intensity $\sigma_{Iobs}$ and the spectral index $\sigma_\alpha$ assuming relative uncertainties of 3\%, 10\%, and 30\%. Similarly, changes in the uncertainty of $\sigma_l$ and $\sigma_{K_O}$ were then examined.

Testing for large (30\%) Gaussian deviations from the central spectral index value $\alpha=0.76$ allows spectral values to become as small as 0.5 or even lower. This excludes the possibility of using the energy distribution approximation as a power law extending to infinity because the integrals, as in Equation (\ref{eq:ep}) (Appendix \ref{sec:cutoff}) become divergent.  Due to the presence of synchrotron and IC losses, which inevitably lead to a high-frequency spectral break, in this analysis we therefore assumed the high-energy cutoff $E_{p2}=E_{e2}=300$\,GeV. This is a reasonable choice given the conditions in the ISM of the Milky Way (see Figure~\ref{fig:energy}) and the MHD simulations of other galaxies \citep{Werhahn2023}.

In this example, we choose a random magnetic field because, at such low LOFAR frequencies, the radio signal from the galactic disk is highly depolarized and does not provide data to model the uniform component of the magnetic field. This allows us to compare the results of the MCMC simulation with the magnetic field value derived from the analytical Equation (\ref{eq:bmag_Br_only}). For the mean values of the observational/model parameters $I_{obs}$, $\alpha$, $l$, and $K_0$ we obtained $B_r=11.82\,\mu$G. Note that if the uncertainties in our model parameters are large, we do not expect the positional estimates of the posterior distribution of $B_r$ to be very close to the analytical value calculated in this way because in the Bayesian approach we use a different data model, with distributions of observational parameters rather than exact values. For the sake of comparison we performed calculations using analytical formulas from \citet{Beck2005} for the case of a random field and obtained $B_r=11.90\,\mu$G. {The difference is due to the assumption $E_{p2}=E_{e2}=\infty$ and the approximate averaging in $c_4$ (Section \ref{sec:simplified}) used in \citet{Beck2005}. According to the BFIELD program accompanying this article, we get $B_r=13.4\,\mu$G. 

\begin{table}[ht]
\caption{Median estimates of the random magnetic field strength for variants of low-frequency fiducial region with different uncertainties of $I_{obs}$, $\alpha$, $l$, and $K_0$.}
\label{tab:median}
\centering
\begin{tabular}{lcccccc}
\hline
\hline
 & \multicolumn{3}{c}{Metropolis-Hastings } & \multicolumn{3}{c}{Affine-Invariant}\\ \hline
 & \multicolumn{3}{c}{$\sigma_{Iobs}$} & \multicolumn{3}{c}{$\sigma_{Iobs}$}\\ 
 \cmidrule(lr){2-4} \cmidrule(lr){5-7}
 & $3\%\, I_{obs}$ & $10\%\, I_{obs}$ & $30\%\, I_{obs}$ & $3\%\, I_{obs}$ & $10\%\, I_{obs}$ & $30\%\, I_{obs}$ \\ 
  \cmidrule(lr){2-4} \cmidrule(lr){5-7}
$B_r$ &  $11.83^{+0.38}_{-0.36}$ & $11.80^{+0.48}_{-0.49}$ & $11.53^{+1.03}_{-1.33}$ & 
$11.83^{+0.36}_{-0.38}$ & $11.81^{+0.48}_{-0.48}$ & $11.54^{+1.03}_{-1.33}$ \\ \hline 

& \multicolumn{3}{c}{$\sigma_{\alpha}$} & \multicolumn{3}{c}{$\sigma_{\alpha}$}\\
 \cmidrule(lr){2-4} \cmidrule(lr){5-7}
& $3\%\, \alpha$ & $10\%\, \alpha$ & $30\%\, \alpha$ & $3\%\, \alpha$ & $10\%\, \alpha$ & $30\%\, \alpha$ \\ 
  \cmidrule(lr){2-4} \cmidrule(lr){5-7}
$B_r$ &  $11.81^{+0.40}_{-0.40}$ &  $11.85^{+0.70}_{-0.58}$ & $12.12^{+3.06}_{-1.28}$ & 
$11.82^{+0.40}_{-0.40}$ & $11.86^{+0.71}_{-0.59}$ & $12.20^{+3.43}_{-1.33}$ \\ \hline 

& \multicolumn{3}{c}{$\sigma_{l}$} & \multicolumn{3}{c}{$\sigma_{l}$}\\ 
 \cmidrule(lr){2-4} \cmidrule(lr){5-7}
& $3\%\, l$ & $10\%\, l$ & $30\%\, l$ & $3\%\, l$ & $10\%\, l$ & $30\%\, l$ \\ 
  \cmidrule(lr){2-4} \cmidrule(lr){5-7}
$B_r$ &  $11.82^{+0.39}_{-0.39}$ & $11.84^{+0.50}_{-0.47}$ & $11.92^{+1.21}_{-0.90}$ & 
$11.82^{+0.38}_{-0.38}$ & $11.84^{+0.50}_{-0.47}$ & $11.93^{+1.30}_{-0.92}$ \\ \hline 

& \multicolumn{3}{c}{$\sigma_{K_0}$} & \multicolumn{3}{c}{$\sigma_{K_0}$}\\ 
 \cmidrule(lr){2-4} \cmidrule(lr){5-7}
& $3\%\, K_0$ & $10\%\, K_0$ & $30\%\, K_0$ & $3\%\, K_0$ & $10\%\, K_0$ & $30\%\, K_0$ \\ 
  \cmidrule(lr){2-4} \cmidrule(lr){5-7}
$B_r$ &  $11.82^{+0.27}_{-0.26}$ & $11.83^{+0.39}_{-0.38}$ & $11.87^{+0.88}_{-1.01}$ & 
$11.82^{+0.26}_{-0.26}$ & $11.81^{+0.39}_{-0.40}$ & $11.88^{+0.87}_{-1.02}$ \\ \hline 
\end{tabular}
\end{table}

For all variations of the low-frequency fiducial region, we ran MCMC M-H simulations for 31 independent chains with 50,000 steps each. The results are shown in Table~\ref{tab:median}. For an uncertainty of 3\% for any of the parameters tested, the relative differences in the median value of $B_r$ are mutually negligible and completely negligible compared to the analytical value, being less than one thousandth. The differences increase to $4\times10^{-3}$ for 10\% uncertainties. The largest relative deviation from the analytical value is only 0.025 (for the 30\% uncertainty of the spectral index). The absolute difference between the median $B_r$ and the analytical value ($0.3\,\mu$G) represents 0.23 of the median uncertainty ($1.28\,\mu$G) and is therefore not statistically significant. Therefore, within the limits of the parameter distributions tested, compared to analytical values, the MCMC methods mainly had the effect of increasing the uncertainty of the determined median value of $B_r$.

We also made an assessment of whether and to what extent the determined magnetic field values depend on the method used for sampling the posterior. Similarly to the analysis above, we used the A-I MCMC sampler to determine the $B_r$ values for identical cases as above running 31 independent chains of 16,000 steps each. The results are shown in the right part of the Table~\ref{tab:median}. Note that unlike the comparison in Section~\ref{sec:EMCEE}, this comparison is made for a different CR energy distribution where the parameters $E_{e2}$ and $E_{p2}$ are finite. However, even in this case the results of the two MCMC methods are very similar. This also confirms that the used MCMC chain lengths are sufficient and suitable to obtain stable results from these simulations.

\subsection{Effect of the synchrotron spectral index}
\label{sec:diff-alpha}

\begin{figure}[ht]
    \centering
    \includegraphics[width=0.49\textwidth]{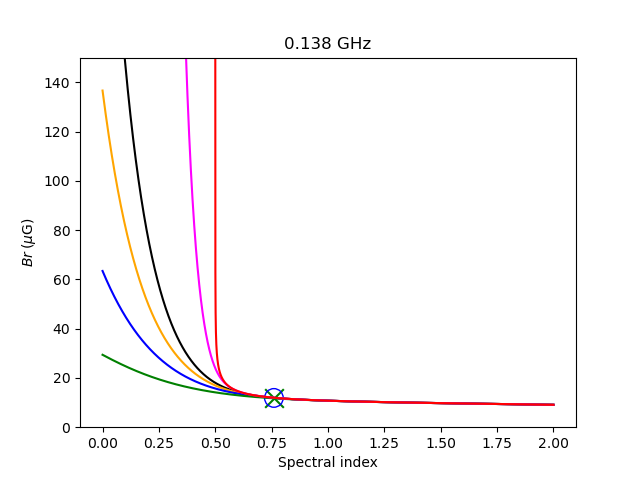}    \includegraphics[width=0.49\textwidth]{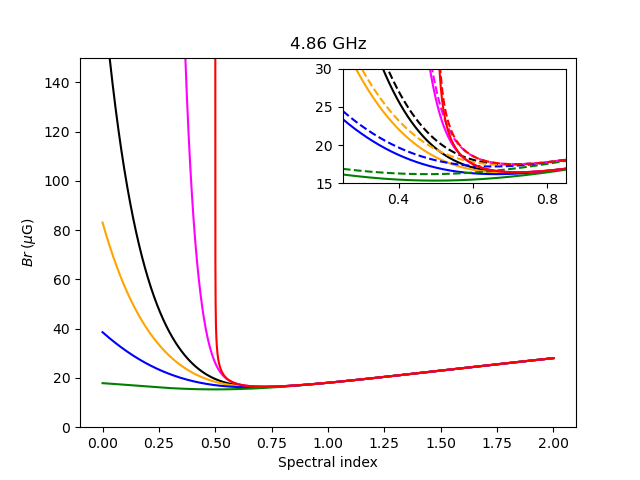}
    \caption{Random magnetic fields derived from the analytical formula versus synchrotron spectral index $\alpha=(\gamma_e -1)/2$, and assuming $\gamma_p=\gamma_e$. Different colors of the solid lines represent various high-energy limits (but the same for CR electrons and protons: $E_{e2}=E_{p2}$) assumed in calculations: green -- $10^{2}$\,GeV; blue -- $10^{3}$\,GeV; orange -- $10^{4}$\,GeV;  black -- $10^{5}$\,GeV; magenta -- $10^{14}$\,GeV; red -- the model assuming no limit ($E_{e2}=E_{p2}=\infty$). Left - $B_r$ for parameters varied from the low-frequency (0.138\,GHz) fiducial region (Section~\ref{sec:comp}). The cross and circle represent arbitrary selected points in the parameter space, which are also represented in Figure~\ref{fig:diff_gamma} by the same symbols. Right -- $B_r$ for parameters varied from the high-frequency (4.86\,GHz) fiducial region (Table~\ref{tab:example_input}). In the upper right corner the smaller inset figure shows the zoomed area of change of the $B_r$ field around $\alpha=0.5$. The dashed lines represent analogous results but from the model of a purely uniform magnetic field $B_u$.}
    \label{fig:spectral_index_dependence}
\end{figure}

In our modeling, the observed synchrotron spectral index determines the CR energy spectrum in the power law part: $\gamma_e=2\alpha +1$. Using the analytical Formula (\ref{eq:bmag_Br_only}) we calculated the equipartition random magnetic field for varying $\alpha$, using other observational/model parameters as in Section~\ref{sec:comp} for the low-frequency fiducial region based on LOFAR observations of NGC\,6946. Then we also performed a series of calculations for different values of high-energy spectral cutoff but the same for protons and electrons ($E_{p2}=E_{e2}$). The results are shown in Figure \ref{fig:spectral_index_dependence} (left panel). Note that the wide range of the $\alpha$ parameter tested in this example is not a kind of magnetic field modeling in NGC\,6946, but rather shows how the field values respond to changes in the parameter space; so each point may actually illustrate a different source or region.

The low observing frequency chosen results in very small changes (up to 1\%) in the value of $B_r$ for $\alpha > 0.7$. In addition, different values of the spectral cutoff ($E_{p2}$, $E_{e2}$) hardly change $B_r$, which is caused by the rapid depletion of high-energy particles for large $\alpha$. The situation is quite different for small $\alpha \le 0.6$. For synchrotron spectral indices close to 0.5, there is a wide range of $B_r$ values for small changes in $\alpha$, both for the model with a spectrum extending to infinity and for models with a finite but very large cutoff energy (e.g. red and magenta dots in Figure \ref{fig:spectral_index_dependence}, respectively). The equipartition model approaches the apparent singularity (for $\gamma=2$) but only for the special case of $E_{p2}=E_{e2}=\infty$ the singularity is real (Appendix \ref{sec:cutoff}).
Although it is theoretically possible for the CR energy spectral index to be less than 2.0 in the high-energy part of the spectrum, it usually requires special conditions such as relativistic shock velocities, specific magnetic field configurations, multiple shocks, or strong nonlinear effects \citep[e.g.][]{Bell1978a, Malkov2001}. Most often, the observed spectra will have slopes higher than 2.0 but for the mathematical completeness in this analysis we show the resulting equipartition magnetic field also for flatter spectra.

The strongly increasing values of $B_r$ for the flat synchrotron spectra are due to the simultaneous flattening of the proton spectra which follows from the assumption $\gamma_p=\gamma_e$. This effect is most pronounced for large values of the high-energy cutoff $E_{p2}$ because then the amount of accumulated protons is highest, causing a strong increase in the equipartition field $B_r$. We obtained similar relationships (Figure \ref{fig:spectral_index_dependence}, right panel) for variants of the high-frequency fiducial region (Table \ref{tab:example_input}) based on observations at 4.86\,GHz. However, in this case, the field strengths increase for steeper synchrotron spectra and are less steep for flat spectra. One of the lines is similar to the plot presented by \citet{Heesen2023} for a model with an unlimited energy spectrum ($E_{p2}=E_{e2}=\infty$). In Appendix \ref{sec:app_figures}, we additionally present families of solutions for models similar to those presented in Figure \ref{fig:spectral_index_dependence}, but modified. We show that increasing the observation frequency increases the dependence of $B_r$ on the spectral slope (Figure \ref{fig:spectral_index_stronger}). Increasing synchrotron intensity values also leads to increased $B_r$ values, but especially for flat spectra. 

We also performed similar modeling, but assuming the purely uniform equipartition magnetic field using Equation (\ref{eq:bmag_Bu_only}) and assuming the field inclination $i=45\degr$ for the reference model (see Table \ref{tab:example_input}). The results are shown as dashed lines in the inset graph of Figure \ref{fig:spectral_index_dependence} (right panel). Similar systematic trends in the effect of the spectral index and high-energy limits of the CR spectra are visible. The differences between the values of the fields $B_r$ and $B_u$ fields are small, less than $1\,\mu$G. However, for larger field inclinations, these differences become larger, e.g. about $5\,\mu$G for $i=60\degr$ and $20\,\mu$G for $i=80\degr$.

\subsection{Different slopes of CR proton and electron energy spectra}
\label{sec:diff_spectra}

\begin{figure}[ht]
   \centering
    \includegraphics[width=0.49\textwidth]{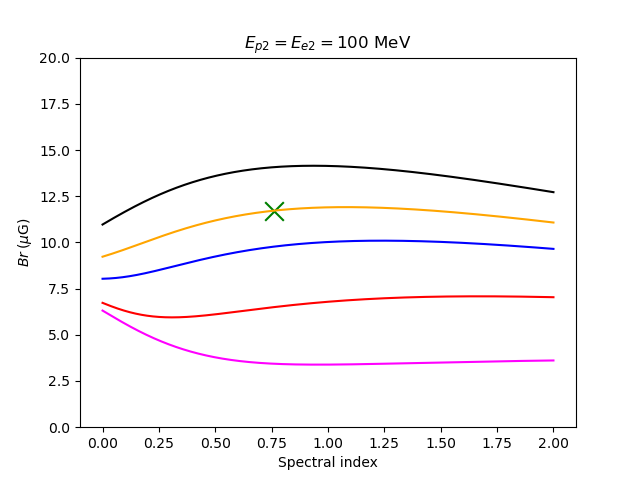}
\includegraphics[width=0.49\textwidth]{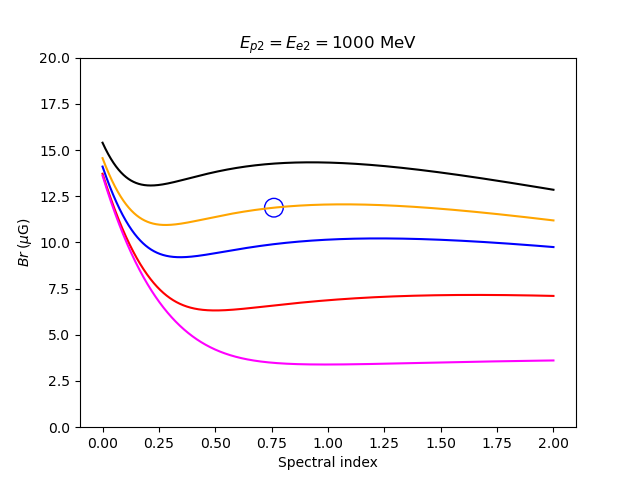}
   \caption{Random magnetic fields derived from analytical formula versus synchrotron spectral index $\alpha=(\gamma_e -1)/2$, and proton energy spectra with $\gamma_p=2.52$. Different colors represent different $K_0$ values: black -- 200; orange -- 100; blue -- 50; red -- 10; magenta -- 0. Left: high energy spectral cutoff $E_{p2}=E_{e2}=100$\,MeV. Right -- $E_{p2}=E_{e2}=1000$\,MeV. The cross and circle represent arbitrary selected points, which are also depicted in Figure~\ref{fig:spectral_index_dependence}. Other parameters correspond to the low-frequency fiducial region.}
    \label{fig:diff_gamma}
\end{figure}

Our equipartition approach allows an analytical calculation of the $B_r$ field even when the energy spectrum of the electrons has a different slope than that of the protons. In particular, an interesting case is shown in Figure \ref{fig:energy}, where the spectrum of electrons measured in the local region of the Milky Way is steeper than that of protons, which may be due to the energy losses of electrons to diffusion, synchrotron radiation and IC scattering. The same is probably true in the regions between the spiral arms and in the outer parts of external galaxies. The results of such modeling are shown in Figure \ref{fig:diff_gamma}, for variants of the low-frequency fiducial region (Section \ref{sec:comp}). It shows the magnetic field values for an assumed constant proton energy spectrum with $\gamma_p=2.52$ but different slopes of the synchrotron spectrum $\alpha$, leading to different $\gamma_e$, and the same high-energy spectral cutoff $E_{p2}=E_{e2}=1000$\,MeV. In addition, a family of solutions is shown for different values of the parameter $K_0$, from zero to 200. The special case of the identical slope $\gamma_p=\gamma_e=2.52$ is marked in Figure \ref{fig:diff_gamma} with a cross symbol, and for comparison, it is also shown identically in Figure \ref{fig:spectral_index_dependence}.

The curve for $K_0=100$ for steep synchrotron spectra shows an almost constant value of the magnetic field independent of the electron spectrum (changes are less than 10\%). For very flat electron spectra, there is a decrease in the value of $B_r$ at a similar small level. This is in contrasts to the dramatic changes in $B_r$ for flat spectra presented in Figure \ref{fig:spectral_index_dependence} and is caused by the assumption of a constant $\gamma_p$ in the current case. For $K_0 = 0$ and $K_0 = 10$ the field strength increases for flat spectra. This reflects the accumulation of energetic electrons and their relatively large effect on the total energy of CRs (for $K_0 = 0$ they are their exclusive component).

On the other hand, for the upper limit of the high-energy cutoff $E_{p2}=E_{e2}=1000$\,MeV shown in Figure \ref{fig:diff_gamma} (right panel), the effect of energetic electrons is even greater, increasing the values of $B_r$ when the spectrum of electrons is flatter than that of protons. In order to complete the presentation of magnetic field changes in the face of different slopes of energy spectra in the appendix Figure \ref{fig:gamma_stronger}, we also present three modifications of the models presented in this section. They show $B_r$ values for 10 times stronger synchrotron emission, 10 times higher observation frequency, and both modifications applied simultaneously. The relationships are no longer flat due to higher frequency, and the field values increase up to about three times.

\subsection{Different cutoffs in CR proton and electron energy spectra}
\label{sec:diff_cutoff}

\begin{figure}
\centering
\begin{minipage}{.49\textwidth} 
  \centering
        \includegraphics[width=1.03\linewidth]{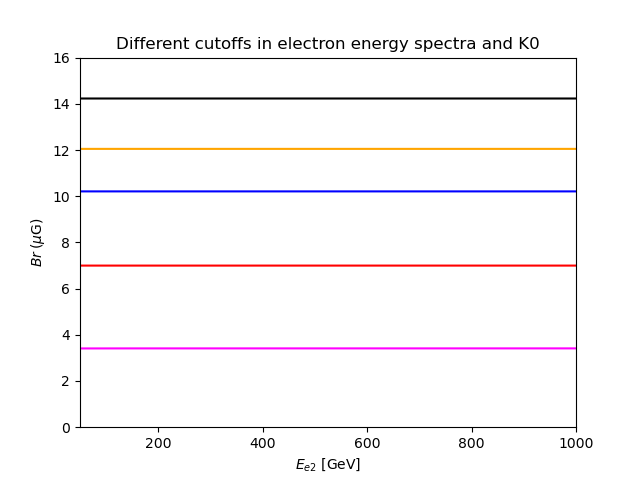} 
    \caption{Random magnetic fields for variants of low-frequency fiducial region with different high-energy cutoffs and $\gamma_p=\gamma_e=2.52$. Different colors represent different $K_0$ values: black -- 200; orange -- 100; blue -- 50; red -- 10; magenta -- 0. The high energy cutoff $E_{p2}=1000$ while $E_{e2}$ is varied.}
  \label{fig:ee2}
\end{minipage}
\begin{minipage}{.49\textwidth}
  \centering
  \includegraphics[width=1.03\linewidth]{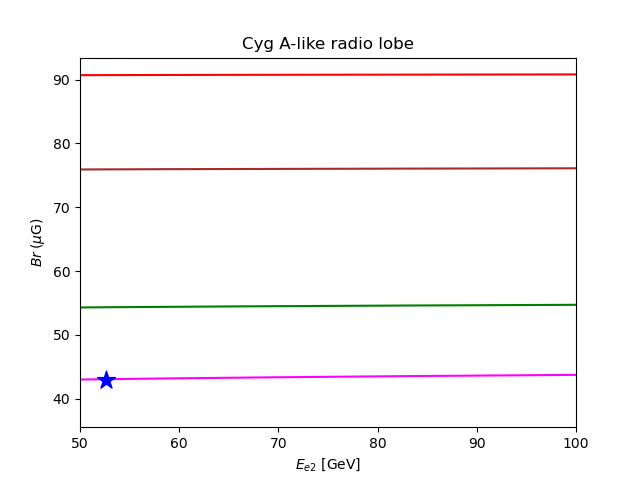}
  \caption{ Similar to Figure \ref{fig:ee2} but for variants of a reference model of a Cyg A-like radio lobe (marked with the asterisk symbol). In this modeling we assumed energy spectra without the flat part and with $\gamma_e=\gamma_p=2.0$. Different colors represent different $K_0$ values: red -- 10; brown -- 5; green -- 1; magenta -- 0.}
  \label{fig:cyga}
\end{minipage}
\end{figure}

Finally, we can ask how much $B_r$ changes when the energy spectrum is constrained from the high-energy end in different ways? This situation may correspond to the energy loss of the CRs which can be approximated here as a spectral cutoff, although in reality it appears more as a steepening or a break. For the electron population, a natural reason for a spectral energy cutoff would be synchrotron or IC cooling. Figure \ref{fig:ee2} gives the answer to this question by showing the modeling of $B_r$ for variations of the low-frequency fiducial region by applying different $E_{e2}$ cutoff values (from 50\,MeV to 1000\,MeV) and keeping a constant cutoff for protons: $E_{p2}=1000$\,GeV. We also assumed a constant value of $\gamma_p=2.52$ and a steeper spectrum for electrons with $\gamma_e=3.3$, similar to the local region in the Milky Way (Section~\ref{sec:energy_spectrum}). As can be seen, over the whole range of variations of the cutoff energy $E_{e2}$, the equipartition magnetic field strengths are virtually identical for a given $K_0$. Thus, under these conditions, the upper limit of the spectrum of protons and electrons has little effect on the magnetic field values. This behavior corresponds to the common path of the curves in Figure \ref{fig:spectral_index_dependence} for typical values of the synchrotron spectrum index, i.e. $\alpha \ge=0.6$). This is also true for plasmas with a large electron population with $K_0=1$ or $K_0=0$. In the latter case, the value of $B_r$ is determined mainly by the energy of the CR electrons in the region close to the breakdown of the flat part of the energy spectrum, where the bulk of particles is located. High energy particles contribute little to the total CR energy and to equipartition with the magnetic field.

Following these results, we examined the behavior of these solutions under conditions where the low-energy part of the CR spectrum lacks a flat part. To illustrate this situation, let us consider the case of a radio galaxy similar to the well-known Cygnus A. To avoid the need for cosmological corrections, we considered such a source but in the closer Universe, at a distance determined by the redshift $z=0.05$. Following the example of the Cyg A lobe presented in the PYSYNCH package of \citep{Hardcastle1998}, we first assumed $K_0=0$ and set a pure power law electron energy spectrum with an index of $\gamma_e=2.0$ and energy limits $E_1=E_e=5.26 \times 10^{-4}$\,GeV and $E_{e2}=52.6$\,GeV, corresponding to the Lorentz factor values in the interval (1, $10^5$). 
We retained from PYSYNCH the physical volume of the lobe of $9.2\times 10^{62}$\,m$^3$ and adjusted the total radio flux at 4.525\,GHz to obtain the equipartition random field of $43\,\mu$G. This value corresponds to the point marked by a triangle in Figure \ref{fig:cyga}.  With this reference model for a Cyg A-like radio lobe, we then varied the parameters $E_{e2}$ and also $K_0$ but only from 0 to 10, since, in contrast to normal galaxies, a less energetically dominant proton population is expected in radio galaxies such as Cyg A \citep{Croston2005}. The results presented in Figure \ref{fig:cyga} show that for a constant $\alpha$ value, even for electron energy spectra without the flat part, the high-energy cutoff does not really affect the magnetic field values. Only a different value of $\alpha$ would alter this relationship. It is worth noting that similar calculations performed in the PYSYNCH package gave similar results.

The magnetic field calculations presented in this and in the previous two sections are based on analytical formulas. The corresponding statistical uncertainties of the determined values of $B$ are not easy to predict, as they nonlineary depend on the specific values of several parameters and their uncertainties. Consequently, the Bayesian modeling method seems to be an appropriate approach to properly estimate them. For all the presented examples it can be easily applied, as shown in Section \ref{sec:example}.

\subsection{BMAG program}
\label{sec:bmag}

We have developed an easy to use web application BMAG (Bayesian MAGnetic field estimation program). It is available at \url{https://bmag.oa.uj.edu.pl}. We used the open-source frameworks Dash and Plotly (under the MIT license) to build an interactive data visualization interface to the actual program written in Python. The code is available upon request. In this program we have implemented the basic features of the Bayesian approach to calculating equipartition magnetic fields presented in this paper. The examples of application of this method reported in Sections~\ref{sec:example}, \ref{sec:EMCEE}, and \ref{sec:comp} are taken entirely from this program.

In the current version of the program, it is possible to:
\begin{itemize}
\item[--] select the topology of the modeled magnetic field: for a purely random or uniform magnetic field the MCMC algorithm is based on analytical Formulas (\ref{eq:Ieq_bmag_Bu}) and (\ref{eq:Ieq_bmag_Br}). For a field composed of uniform plus random fields Equations (\ref{eq:Ieq_bmag}) and (\ref{eq:Peq_bmag}) are used (in this case, the computation time is considerably longer in time due to execution of the numerical integrations);
\item[--] set a diffuse prior for magnetic field distributions;
\item[--] use a Gaussian model for the uncertainties of the data parameters $I_{obs}$, $PI_{obs}$, $\alpha$, $l$;
\item[--] choose a prior model for the parameter $K_0$ as a constant or a Gaussian distribution;
\item[--] specify the high-energy spectra for CR electrons and protons as a power law extending to infinity or with a finite cutoff energy, assuming $E_{p2}=E_{e2}$;
\item[--] set the low-energy spectral break $E_p$, assuming $E_p=E_e$;
\item[--] choose the sampling method of the posterior distribution: Metropolis-Hastings or Affine-Invariant sampler;
\item[--] set simulation parameters such as number of chains, burn-in steps, and total number of steps in chains. For the M-H method, the user can specify additional parameters of the Gaussian proposal function for $I_{obs}$, $PI_{obs}$, $\alpha$, $l$, and $K_0$.
\end{itemize}

The Dash web interface runs in a separate thread. The computation starts with a search for the global maximum of the posterior in a multidimensional parameter space using the stochastic differential evolution method from the SciPy package. The initial values of the chains for each model parameter are then found as a Gaussian distribution around the posterior maximum. The main posterior sampling is then performed using one of the two methods selected by the user. The massive Monte Carlo computations are parallelized, so that the walkers are computed on different processor cores. The program returns diagnostic figures that allow for easy assessment of the quality of the simulation and convergence to the target distribution (see Figures \ref{fig:mh_traceplot}, \ref{fig:mh_correl}). The determined magnetic field values from the marginalized posterior distribution are presented in the form of three position measures (mean, median, mode) with calculated uncertainties at a level corresponding to $1\sigma$ (68\%), as well as a corner plot (see Table \ref{tab:mh_estimates} and Figure \ref{fig:mh_corner}). In addition, for uniform only or random only magnetic fields, the equipartition field values from analytical Equations (\ref{eq:bmag_Bu_only}) and (\ref{eq:bmag_Br_only}) are also given for comparison. The compressed archive containing all files with input parameters, output results, and figures is available to download by the user as soon as the calculations are finished.

\subsection{Recommendations}
\label{sec:recommendations}

Based on our analyses in Sections \ref{sec:EMCEE} - \ref{sec:diff_cutoff}, we are tempted to summarize them in the form of brief recommendations on which parameters to use in the equipartition method and, in the absence of sufficient information (which is most often the case), how to use values consistent with previous studies. First of all, it should be noted that the formulas presented for the equipartition magnetic field require that the observed intensity of synchrotron emission be specified. Therefore, the separating of the nonthermal emission from the thermal free-free radiation is a necessary preliminary step. At low frequencies (e.g., LOFAR), the thermal contribution can be neglected in most cases. However, at low frequencies in a dense interstellar environment, radiation absorption may become prominent and should then be modeled accordingly, since the equipartition formulas assume optically thin synchrotron emission.

We also note, as mentioned in Section~\ref{sec:field_configuration}, that since the synchrotron polarization depends only on the orientation of the magnetic field and not on the direction, the $B_u$ value determined from the equipartition method can represent a regular field or a large-scale field with direction reversals. $B_u$ can also be influenced or even dominated by anisotropic random fields, which can arise from an isotropic random component in the process of compressing or shearing gas flows. All these types of fields leading to net polarization are called ordered fields \citep{Beck2015}. Observational information on the total and linearly polarized synchrotron intensities is insufficient to unambiguously determine the magnetic-field topology. Additional information is needed to identify the type of field present in the observed source and to accurately interpret the observed polarized emission \citep[see e.g.][]{Fletcher2011, Muller2020, Paraschos2024}.

Theoretical considerations by \citet{Bell1978b} show that for a typical galactic supernova shock front moving at 10,000\,km\,s$^{-1}$, the acceleration of CRs results in a proton/electron ratio of about 100 at energies above of 1\,GeV. This value is similar to that measured in situ near the Solar System (Section \ref{sec:energy_spectrum}). Thus, for normal galaxies, it is reasonable to use the constant $K_0=100$ (see also Section \ref{sec:introduction}), unless we know the ISM parameters, the particle acceleration and the propagation processes in the studied galactic region. The parameter $K_0$ should also include the population of primary positrons and secondary $e^- + e^+$, but their expected contribution is rather small (up to 30\% in energy) as shown in simulations of starbursts by \citet{Pfrommer2022}. Lower values of $K_0$ seem to be appropriate for active galaxies. In the papers of \citet{Hardcastle2002} and \citet{Croston2005}, among others (and in our modeling of magnetic field values in Figure \ref{fig:cyga}), even $K_0=0$ was used, corresponding corresponds to a purely leptonic plasma.

As shown in Section \ref{sec:diff-alpha} (see Figure \ref{fig:spectral_index_dependence}), the values of the magnetic field strengths depend very strongly on the index of synchrotron radiation when the spectra are flat $\alpha \le 0.6$ and "banana-like" relationships between the parameters then arise. In this range of the spectral index, it is advantageous to have very good estimates of it, since narrowing the prior distributions for $\alpha$ significantly reduces the uncertainties in the magnetic field estimates. The assumption $E_{p2}=E_{e2}=\infty$ is commonly used in equipartition method but can only be applied for large values of observed synchrotron spectral index and small uncertainties. The use of this approximation for small values of $\alpha$ is nonphysical as it leads to infinite energy of CR electrons (Appendix \ref{sec:cutoff}). In this case, it is appropriate to adopt a different energy distribution, with a finite value of $E_{e2}$. 

The use of a high-frequency radio spectral index, to determine the magnetic field from the equipartition assumption may not be correct in the case of strong curvature in synchrotron spectra, which may be due to energy losses of electrons to radiation and IC scattering. In this case, the magnetic field values derived by the equipartition approach with constant $\alpha$ overestimate the true values. One would then need to use the fitted curved electron energy spectrum to correctly determine the magnetic field (beyond scope of this paper, although possible with our Bayesian approach). A possible solution to this problem may be to use the low-frequency (e.g. from LOFAR) spectral index $\alpha$ as a better approximation of the spectra of protons ($\gamma_p=2 \alpha + 1$) and electrons without high-energy losses.
As explained in Appendix A of \cite{Heesen2023} discussing the \cite{Beck2005} approach to equipartition, using a $K_0$ value greater than 100 and a steep spectral index $\gamma$, in an attempt to account for CR electron energy losses, leads to a strong overestimation of the equipartition magnetic field. On the other hand, using a fixed $K_0$ and a flat $\gamma$, as in the inner region of the galaxy, actually leads to an underestimation of the equipartition field, since the synchrotron intensity is reduced due to the lower CR electron density. The solution may be to use a $K_0$ value greater than 100 and a flat $\gamma$, as in the inner regions of the galaxy. This suggestion is missing in \cite{Heesen2023}.

Our analysis shows that the low-energy spectral break for protons $E_p$ also strongly affects the estimated values of the magnetic field. From the measurements of the Milky Way CRs (Figure \ref{fig:energy}) and modeling of nearby starbursts \citep{Werhahn2021b}, it appears that the spectral break for protons occurs near the rest mass of the proton and this value can be used as a reasonable choice for $E_p$ in galaxies. When $K_0=0$ (leptonic plasma), the spectral break for electrons $E_e$ can be set similar to the rest mass of the electron. In contrast, the specific value of the high-energy spectral cutoff for steep spectra is practically irrelevant for equipartition magnetic fields (see Figures \ref{fig:ee2} and \ref{fig:cyga}).

To determine the magnetic field strength from synchrotron emission, one can use observational data other than those used in Section \ref{sec:example}. For example, the average magnetic field can be calculated for a larger spatial region, a whole galaxy, or a lobe in a radio galaxy. In this case, the  corresponding integrated flux can be used in Formula (\ref{eq:KS18_intensity}) and the following, and the angular size of the source should substitute the beam size ($\theta_{maj}$, $\theta_{min}$). It is also necessary to remove the factor of $\pi / ({4\ln 2)}$ that was used in the case of flux per beam. The example shown in Figure \ref{fig:cyga} was calculated in this way.

When using BMAG to perform MCMC simulations, attention should be paid to the correctness of the simulation results obtained. First, if after the declared burn-in period the simulated chains in the diagnostic drawings (e.g. Figure \ref{fig:mh_traceplot} and \ref{fig:em_traceplot}) do not overlap and stabilize within a common spread in the declared burn-in period, a longer burn-in period is required. Similarly, if the rank normalized $\hat{R}$ diagnostic statistic exceeds the recommended threshold of 1.01 \citep{Vehtari2021} then there is a concern that the chains have not converged, so they may not produce representative samples from the target distribution. In this case, the length of the chains should be increased accordingly. In the case of the M-H method the size of the steps in sampling the posterior distribution can also be adjusted.

\section{Summary}
\label{sec:summary}

In this paper we present a new approach to calculate the magnetic field strength from energy equipartition between CRs and magnetic fields. We used a Bayesian approach in which the magnetic field is treated as the posterior distribution, which is calculated directly from the synchrotron theory formulas used in the likelihood function. Thus, we have avoided the need to reverse these formulas, as has been done in the past and have limited the cases in which the equipartition method can be applied. 

Unlike previous approaches, we have derived a formalism that allows us to apply the equipartition method in the general case of a magnetic field consisting of both a uniform component and randomly oriented field component of the constant strength. We have also extended the previous concepts and derived the equipartition formula for different energy distributions of CR electrons and protons, i.e. for different slopes of their power spectra, different breaks at low energies and different cutoff energies at the high-energy end. Such spectra are measured in situ in the local region of the Milky Way and appear in galaxy simulations. We show that the formalism can also be applied to a purely uniform or random field, and in these simplified cases we obtained analytical formulas for the magnetic field strength. Another advantage of the applied Bayesian approach is that it automatically provides from the posterior distribution the uncertainties in the estimated magnetic field strength, which result from the uncertainties in the observables and the assumed values of the unknown physical parameters.

We used MCMC methods to obtain the posterior distribution of the magnetic field strength. In the example calculations, we applied two independent simulation codes: Metropolis-Hastings and Affine-Invariant (from the EMCEE package) to demonstrate that our results are not biased by the computational approach. Diagnostic plots and the rank-normalized $\hat{R}$ statistics were used to ensure that the simulated chains achieved stable results and converged to the target distribution. The resulting distributions of the total magnetic field and the regular and random components were presented using different point estimators (mean, median, and mode) and credible intervals at the 68\% probability level.

Using analytical formulas to determine the equipartition magnetic field in the case of a random field, we showed how different values in parameter space affect the field, how they are correlated, and gave recommendations on how to use the method. We presented families of solutions for magnetic fields as a function of the synchrotron spectral index and observation frequency. At low frequencies the magnetic field is weakly dependent on the spectral index. However, the shape of this dependence is not universal and the dependence slowly increases with the frequency of observations and the intensity of the nonthermal emission. The magnetic field values become very strongly dependent on the synchrotron spectral index when the spectra are flat $\alpha \le 0.6$. In this range, it is advantageous to have very good (prior) estimates of $\alpha$ to minimize the uncertainties in the magnetic field estimates. We show that the commonly used assumption $E_{p2}=E_{e2}=\infty$ is unphysical in this spectral range as it leads to an infinite energy of CRs, and it should be replaced by finite values of cutoffs. Also inappropriate is the use of the high-frequency synchrotron spectral index when radiative or IC energy losses are significant: in this case, the use of low-frequency radio data or a thorough modeling of the energy spectra (beyond the scope of this work) is recommended. In contrast, the specific value of the high-energy spectral cutoff for steep synchrotron spectra is practically irrelevant for equipartition magnetic fields, since high-energy particles contribute little to the total CR energy budget. We also used parameter values suitable for radio galaxies, while dropping the flattened part of the spectrum in favor of the low-energy cutoff. Again, regardless of the value of $K_0$, the specific high-energy cutoff value has little effect on the magnetic field values.

We have developed a web application BMAG (\url{https://bmag.oa.uj.edu.pl}) that applies the presented approach under real astrophysical conditions, making our method adaptable to different values of model and observational parameters of real sources.
It performs M-H and A-I MCMC simulations to sample posterior magnetic fields and presents them as shown in this paper. For comparison, where possible, it also gives results from analytical formulas (for magnetic fields containing only a uniform or random component).

This work has presented simple examples of the application of the Bayesian approach to equipartition magnetic fields. However, the proposed method gives the possibility to apply equipartition to more complex, curved CR energy spectra, which is appropriate in the case of strong energy loss effects of CR particles. Furthermore, a straightforward development of this Bayesian approach can lead to the construction of magnetic maps over the sources together with field uncertainty maps, which should greatly facilitate the interpretation of magnetic fields in various objects \citep[see e.g.][]{ChyzyButa2008}. However, using a mixture of both field components (uniform and random) would significantly lengthen MCMC computations over time and require fast computer clusters.

Future multiband observations (e.g. with the SKA and LOFAR) will provide detailed radio frequency spectra, and thus allow for the CR energy losses observed at low and high RF frequencies to be adequately taken into account in the equipartition method. Together with X-ray and gamma-ray telescopes, this will also enable accurate modeling of CR evolution in galaxies, which should resolve and verify the equipartition paradigm. 

\begin{acknowledgments}

Acknowledgments. We thank Rainer Beck, Marek We\.zgowiec and Marek Jamrozy for their valuable comments on the manuscript. We also thank Aleksander Kurek for comments on the BMAG code. The authors express their gratitude to the anonymous referee for helpful comments that improved the quality of the manuscript. This research made use of matplotlib -- a Python library for publication-quality graphics \citep{Hunter:2007}; SciPy \citep{Virtanen_2020}; NumPy \citep{harris2020array}; ArViz \citep{arviz}, Wolfram Mathematica \citep{Mathematica}, services provided by the Space Science Data Center (SSDC); Cosmic-Ray Data Base \citep[CRDB,][]{Maurin2023}.

We have used data from the International LOFAR Telescope (ILT). LOFAR \citep{vanHaarlem2013} is the Low Frequency Array designed and constructed by ASTRON. It has observing, data processing, and data storage facilities in several countries, that are owned by various parties (each with their own funding sources), and that are collectively operated by the ILT foundation under a joint scientific policy. The ILT resources have benefited from the following recent major funding sources: CNRS-INSU, Observatoire de Paris and Université d'Orléans, France; BMBF, MIWF-NRW, MPG, Germany; Science Foundation Ireland (SFI), Department of Business, Enterprise and Innovation (DBEI), Ireland; NWO, The Netherlands; The Science and Technology Facilities Council, UK; Ministry of Science and Higher Education (MSHE), Poland. We also thank MSHE for providing funding for the Polish contribution to the ILT (MSHE decision no. DIR/WK/2016/2017/05-1) and for the maintenance of the stations LOFAR PL-610 Borowiec, LOFAR PL-611 Lazy, LOFAR PL-612 Baldy.
\end{acknowledgments}

\appendix
\section{Simplified CR energy spectra}
\label{sec:Appendix_special_cases}

Energy spectra of CRs postulated in Equations (\ref{eq:np}) and (\ref{eq:ne}) allow for great freedom in manipulating the shape and relative number density of protons and electrons. Including the assumption of energy equipartition provides a constraint crucial for deriving a closed-form expression describing the constant in electron density $N_e$ given by Equation (\ref{eq:N_e}). This normalization later leads to the total and polarized intensities given by Equations (\ref{eq:Ieq}) and (\ref{eq:Peq}). These formulas allow full flexibility in the choice of parameters for the modeled CR energy spectrum. In this appendix, we discuss selected special cases and simplifications that can be made by specific choices of numerous parameters appearing in Equation (\ref{eq:N_e}).

\subsection{Similar shape of the energy spectra of electrons and protons}

If we consider the energy spectra of electrons and protons to be of similar shape, we can put $\gamma_p=\gamma_e\equiv\gamma$, which means that we consider the energy spectra of electrons and protons that have the same slope. Similarly, we can choose the same low-energy break $E_p=E_e$ and high-energy cutoff $E_{e2}=E_{p2} \equiv E_2$. Except for finite $E_2<\infty$, this is the type of spectrum visualized in the right part of Figure \ref{fig:model_energy}. This choice of energy spectrum leads to the synchrotron total and polarized intensity given by:

\begin{equation}
\label{eq:Ieq_bmag}
    I_\nu= \frac
      {
      \frac{B_{eq}^2}{8 \pi} \left(\frac{\nu}{2c_1}\right)^\frac{1-\gamma}{2} \frac{c_2 f l}{4 \pi} \int_{4\pi}(B\sin{\mu})^{\frac{\gamma+1}{2}}d\Omega
      }
      {(K_0+1)   \frac{E_p^{2-\gamma}\gamma}{2(\gamma-2)}\left(1- \frac{2}{\gamma}\left(\frac{E_{2}}{E_p}\right)^{2-\gamma}\right)
      },
\end{equation}

\begin{equation}
\label{eq:Peq_bmag} 
    {PI}_\nu=\frac
    {
    \frac{B_{eq}^2}{8 \pi} \left(\frac{\nu}{2c_1}\right)^\frac{1-\gamma}{2} \frac{p_0 c_2 f l}{4 \pi} \int_{4\pi}(B\sin{\mu})^{\frac{\gamma+1}{2}} \cos{2\chi}d\Omega
    }
    {
    (K_0+1) \frac{E_p^{2-\gamma}\gamma}{2(\gamma-2)}\left(1- \frac{2}{\gamma}\left(\frac{E_{2}}{E_p}\right)^{2-\gamma}\right)
    }.
\end{equation}

Formulas (\ref{eq:Ieq_bmag}) and (\ref{eq:Peq_bmag}) are used to calculate the synchrotron intensity by the BMAG program described in Section \ref{sec:bmag}.

\subsection{Special cases of magnetic field}
As it was explained in Section \ref{sec:special_cases}, for special cases of either purely uniform or purely random magnetic fields, the formulas describing the synchrotron intensity are further simplified and can be inverted to recover the value of the magnetic field strength. For a purely uniform field, Equation (\ref{eq:Ieq_bmag}) simplifies to:

\begin{equation}
\label{eq:Ieq_bmag_Bu}
    I_\nu= \frac
      {
      \frac{B_{u}^2}{8 \pi} \left(\frac{\nu}{2c_1}\right)^\frac{1-\gamma}{2} c_2 f l \left(B_{u} \sin{\theta_0}\right)^{(\gamma+1)/2}
      }
      {(K_0+1)   \frac{E_p^{2-\gamma}\gamma}{2(\gamma-2)}\left(1- \frac{2}{\gamma}\left(\frac{E_{2}}{E_p}\right)^{2-\gamma}\right)
      },
\end{equation}
which can be solved for $B_{u}$:

\begin{equation}
\label{eq:bmag_Bu_only}
B_{u}=
\biggl\{
    \frac{
        8\pi \left(\frac{\nu}{2c_1}\right)^\frac{\gamma-1}{2}I_\nu (K_0+1)  
    }
    {
        \left(\sin{\theta_0}\right)^{\frac{\gamma+1}{2}} c_2 f l
    } 
    \frac{E_p^{2-\gamma}\gamma}{2(\gamma-2)}\left(1- \frac{2}{\gamma}\left(\frac{E_{2}}{E_p}\right)^{2-\gamma}\right)
\biggr\}^{\frac{2}{\gamma+5}}.
\end{equation}
Similarly, for the purely random field, Equation (\ref{eq:Ieq_bmag}) gives:

\begin{equation}
\label{eq:Ieq_bmag_Br}
    I_\nu= \frac
      {
      \frac{B_{r}^2}{8 \pi} \left(\frac{\nu}{2c_1}\right)^\frac{1-\gamma}{2} c_2 f l B_{r}^{\frac{\gamma+1}{2}} \frac{\sqrt{\pi}}{2} \frac{\Gamma\left(\frac{\gamma+5}{4}\right)}{\Gamma\left(\frac{\gamma+7}{4}\right)}
      }
      {
      (K_0+1)   \frac{E_p^{2-\gamma}\gamma}{2(\gamma-2)}\left(1- \frac{2}{\gamma}\left(\frac{E_{2}}{E_p}\right)^{2-\gamma}\right)
      },
\end{equation}
which leads to the expression for $B_r$:

\begin{equation}
\label{eq:bmag_Br_only}
B_{r}=
\biggl\{
    \frac{8\pi \left(\frac{\nu}{2c_1}\right)^\frac{\gamma-1}{2} I_\nu (K_0+1)}{ c_2 f l} 
    \frac{E_p^{2-\gamma}\gamma} {2 \left( \gamma - 2 \right) } \left(1- \frac{2}{\gamma}\left(\frac{E_2}{E_{p}}\right)^{2-\gamma}\right) 
    \frac{2}{\sqrt{\pi}} \frac{\Gamma\left(\frac{\gamma+7}{4}\right)}{\Gamma\left(\frac{\gamma+5}{4}\right)} 
\biggr\}^{\frac{2}{\gamma+5}}.
\end{equation}

Equations (\ref{eq:Ieq_bmag_Bu}) and (\ref{eq:Ieq_bmag_Br}) together with $\gamma=2\alpha+1$ are used by the BMAG program for the MCMC simulation of a purely uniform or purely random field, respectively, while Equations (\ref{eq:bmag_Bu_only}) and (\ref{eq:bmag_Br_only}) are used for analytical values of magnetic fields also calculated by the program.

\subsection{Note on apparent singularity}
Equation (\ref{eq:N_e}), and consequently all radiation intensity formulas such as Equations (\ref{eq:Ieq}) and (\ref{eq:Peq}), appear to be singular at $\gamma_e=2$ due to the result of integrating the total energy density in the power law part of the energy spectrum:

\begin{equation}
    \int_{E_e}^{E_{e2}}  N_e E^{-\gamma_e} E dE=N_e\frac{E_e^{2-\gamma_e} -E_{e2}^{2-\gamma_e}}{\gamma_e-2}.
\end{equation}
This formula is therefore only valid for $\gamma\neq2$. For $\gamma=2$ the above integral has the following solution:
\begin{equation}
    \int_{E_e}^{E_{e2}}  N_e E^{-\gamma_e} E dE\,\overset{\gamma=2}{=}\int_{E_e}^{E_{e2}}  N_e E^{-1}dE=N_e\ln{\left(\frac{E_{e2}}{E_e}\right)}.
\end{equation}
The same reasoning applies to protons. Taking this into account all resulting quantities remain smooth when passing through $\gamma_e=2$ and all formulas remain valid arbitrarily close to $\gamma_e=2$, as demonstrated, for example, in Figure \ref{fig:spectral_index_dependence}. Nevertheless, greater caution is advised in proper handling of numerical calculations in this range of values of $\gamma_e$ or~$\gamma_p$.

\subsection{High energy cutoff}
\label{sec:cutoff}

As shown in Figure \ref{fig:spectral_index_dependence}, for values of the synchrotron spectral index sufficiently higher than $\alpha=0.5$ (corresponding to $\gamma=2$) the effect of choosing a specific value of the high-energy cutoff $E_{2}$ is negligible. For $\gamma>2$ factors of the form $\left(E_{2} / E_p \right)^{2-\gamma}$ like in Equations (\ref{eq:bmag_Bu_only}) and (\ref{eq:bmag_Br_only}) quickly approach 0 as $E_{2}$ approaches infinity. This simplification introduces the true singularity of the resulting formulas at $\gamma=2$. In this case, the prior for the synchrotron spectral index should be limited to values above $\alpha > \alpha_{min}=0.5$  (cf. Section~\ref{sec:prior}).

\subsection{Comparison with previous approaches}
\label{sec:simplified}
When we assume that the energy spectra extend to infinity ($E_{p2}=E_{e2}=\infty$), and use the synchrotron index $\alpha$ instead of the energy index $\gamma$, then Equation (\ref{eq:bmag_Bu_only}) for the uniform magnetic field is simplified to:
\begin{equation}
\label{eq:Beq_uni_Beck}
B_{u}=\biggl\{4\pi  (K_0+1)  \frac{2\alpha+1}{2\alpha-1} \left(\frac{\nu}{2c_1}\right)^{\alpha} \frac{I_\nu E_p^{1-2\alpha}}{c_2 l} (\sin \theta_0)^{-(\alpha+1)}\biggr\}^{\frac{1}{\alpha+3}}.
\end{equation}
This is exactly Equation (A18) from \citet{Beck2005}, which shows that for the case of a purely uniform magnetic field and similar energy spectra of protons and electrons, our approach reduces to the previously known one. Analogous treatment of the purely random magnetic fields (Equation \ref{eq:bmag_Br_only}) but without conversion to the $\alpha$ parameter leads to:

\begin{equation}
\label{eq:Beq_ran_g}
B_{r}=\biggl\{4\pi \left(K_0+1\right)  \frac{\gamma}{(\gamma-2)} \left(\frac{\nu}{2c_1}\right)^{\frac{\gamma-1}{2}}\frac{I_\nu E_p^{2-\gamma}}{c_2 l} \frac{2}{\sqrt{\pi}} \frac{\Gamma\left(\frac{\gamma+7}{4}\right)}{\Gamma\left(\frac{\gamma+5}{4}\right)} \biggr\}^{\frac{2}{\gamma+5}}.
\end{equation}

The effect of averaging of the projected magnetic field on the sky plane is described in \citet{Beck2005} by the symbol $c_4$ dependent on field inclination $i$:
\begin{equation}
    B_\perp^{\frac{\gamma+1}{2}}=B^{\frac{\gamma+1}{2}}c_4(i).
\end{equation}
For example, for a purely uniform field $c_4=c_4(i)=\cos(i)^{(\gamma+1)/2}$, which corresponds to $c_4(\theta_0)=\sin(\theta_0)^{(\gamma+1)/2}$ for the angle naming conventions used in the derivation of the formulas in this paper. For a random magnetic field we obtain:
\begin{equation}
\label{eq:c4_by_me}
c_4=\frac{\sqrt{\pi}}{2}\frac{\Gamma\left(\frac{\gamma+5}{4}\right)}{\Gamma\left(\frac{\gamma+7}{4}\right)}.
\end{equation}
A different averaging approach was used by \citet{Beck2005} where the square of the projected magnetic field in the plane of the sky was first averaged, giving $<B_\perp^2>=(2/3)B^2$, and then raised to the power of $(\gamma+1)/4$, giving $c_4=(2/3)^{(\gamma+1)/4}$. The effect of this approximation on the final values of the estimated $B_r$ is shown in Figure \ref{fig:c4} together with the results of the exact averaging of $B_\perp^{(\gamma+1)/2}$, which leads to Equation (\ref{eq:c4_by_me}). The differences are small, generally in the order of 1\%--3\%, and vanish for $\gamma = 3$. Although simpler averaging gives quite accurate values of $B_r$, in this work we use strict formulas to average the field orientation in the case of a purely random magnetic field.

\begin{figure}[ht!]
    \centering
    \includegraphics[width=0.9\textwidth]{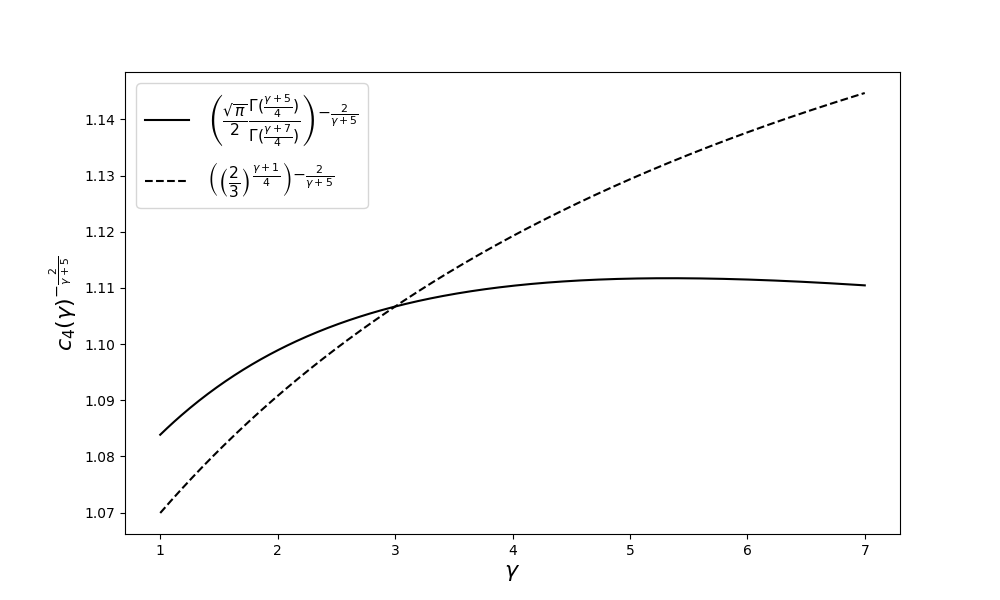}
    \caption{Comparison of random magnetic field values obtained by the exact and approximate field averaging leading to different values of $c_4$ and hence different values of $B_{r}\sim c_4 ^{-2/(\gamma+5)}$. Solid line: exact averaging leading to $c_4$ given by Equation~(\ref{eq:c4_by_me}). Dashed line: simpler averaging used by \citet{Beck2005}.} 
    \label{fig:c4}
\end{figure}

\section{Alternative choice of prior distribution}
\label{sec:app_prior}

In this work, we used a uniform prior on the values of $B_u$ and $B_r$ (see Section \ref{sec:prior}). Another well-motivated choice of non-informative prior for $B_u$ and $B_r$ could be the Jeffreys prior \citep{JP}, which is invariant under parameter transformations. However, the implementation of the Jeffreys prior for magnetic fields containing both uniform and random components would require the determination of the Fisher information matrix and hence the computation of partial derivatives of the integral expressions in Equations (\ref{eq:Ieq}) and (\ref{eq:Peq}) for synchrotron emission. This would result in several additional numerical integrations at each step of the MCMC simulation and multiple increase in the computational time compared to using a uniform prior.

We tested the choice of the Jeffreys prior in the simple cases of pure $B_u$ and pure $B_r$, using Equations (\ref{eq:Ieq_bmag_Bu}) and (\ref{eq:Ieq_bmag_Br}), respectively, which are analytical and do not require numerical integrations. We compared the differences between the values of the mean, median, and mode of the magnetic field obtained from MCMC simulations using the Jeffreys and uniform priors. We found that for each of these parameters, the differences are only on the order of 1\%--2\% of their uncertainties given by 68\% credible intervals. Therefore, we conclude that in these cases the Jeffreys prior does not cause significant differences from the uniform prior.

\section{Additional figures}
\label{sec:app_figures}

We used an alternative Bayesian method to determine magnetic field strengths based on the Affine-Invariant MCMC algorithm for sampling the posterior distributions of the magnetic fields (see Section \ref{sec:EMCEE}). The resulting corner plot (Figure \ref{fig:em_corner}) and diagnostic plots (Figures \ref{fig:em_traceplot} and \ref{fig:em_correl}) are presented below.

We also display results (Figures \ref{fig:spectral_index_stronger} and \ref{fig:gamma_stronger}) of an additional analysis to that presented in Sections \ref{sec:diff-alpha} and \ref{sec:diff_spectra}, using analytical solutions for the equipartition random magnetic field. They show the change in the dependence of $B_r$ on the synchrotron spectral index when a different synchrotron intensity or observation frequency is applied.

\begin{figure}[h]
\plotone{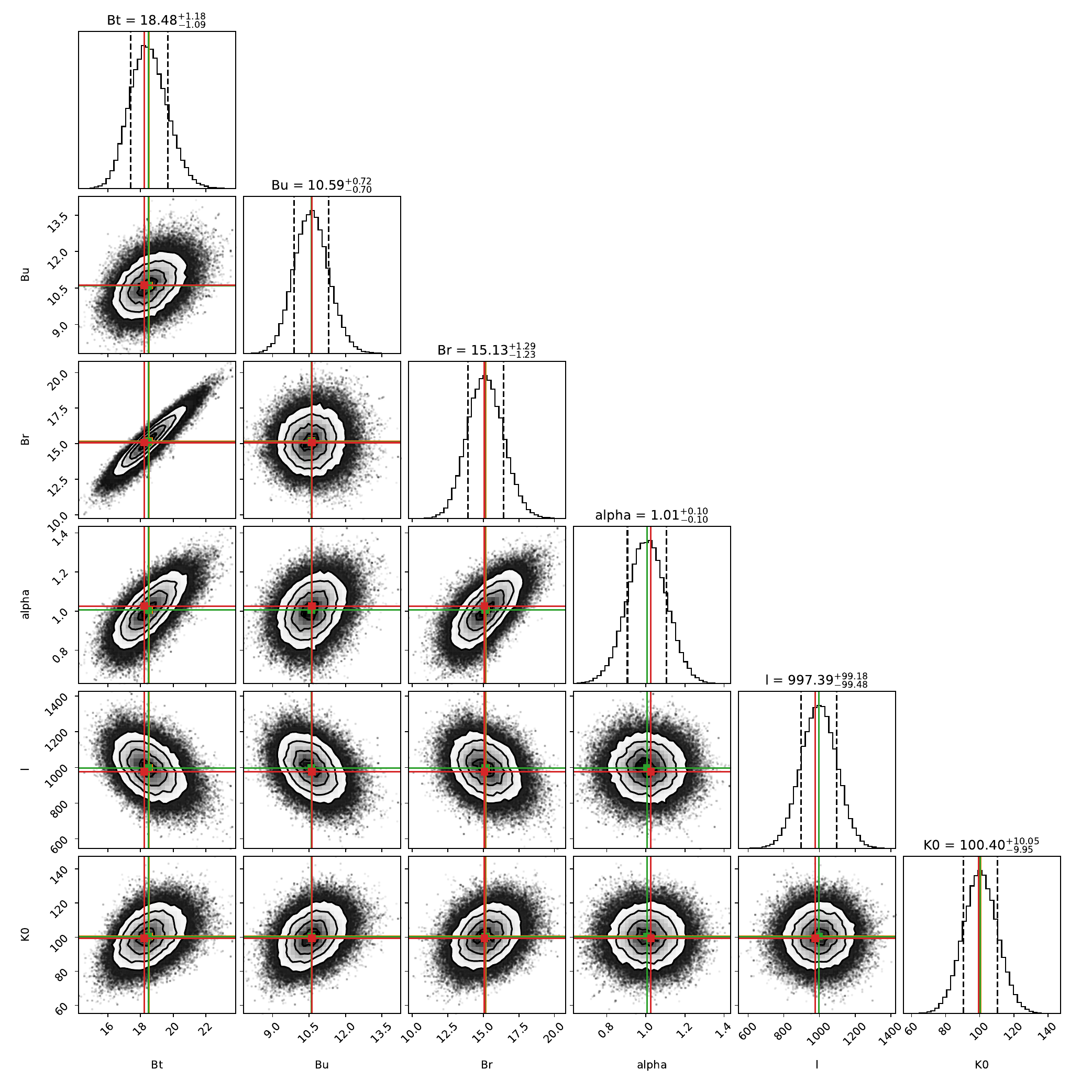}
\caption{Corner plot from the MCMC simulations using the A-I sampler showing the posterior distributions of the magnetic field components. See Figure \ref{fig:mh_corner} for a comparison.} 
\label{fig:em_corner}
\end{figure}

\begin{figure}[h]
\begin{center}
\includegraphics[width=8.9cm]{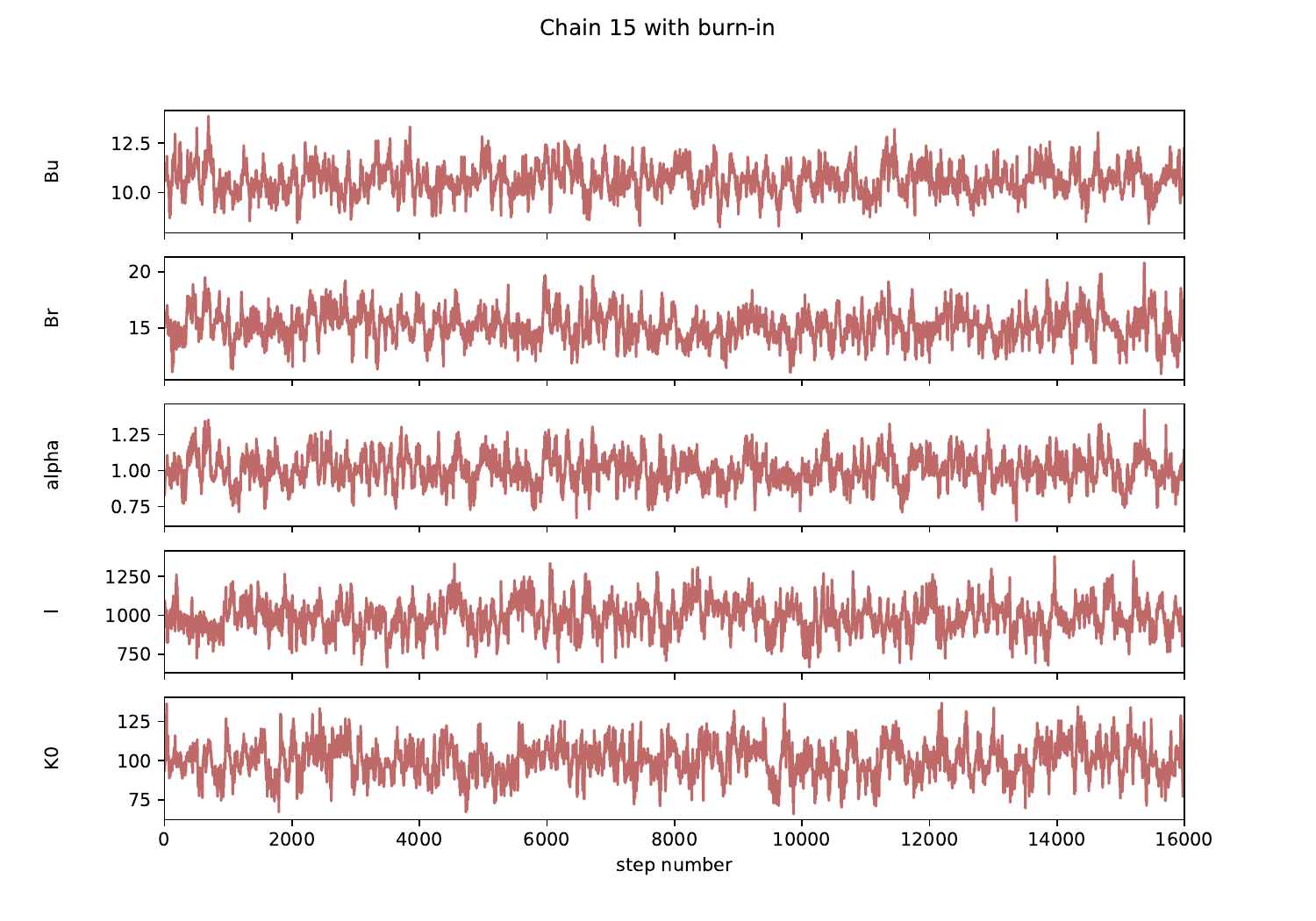}
\includegraphics[width=8.9cm]{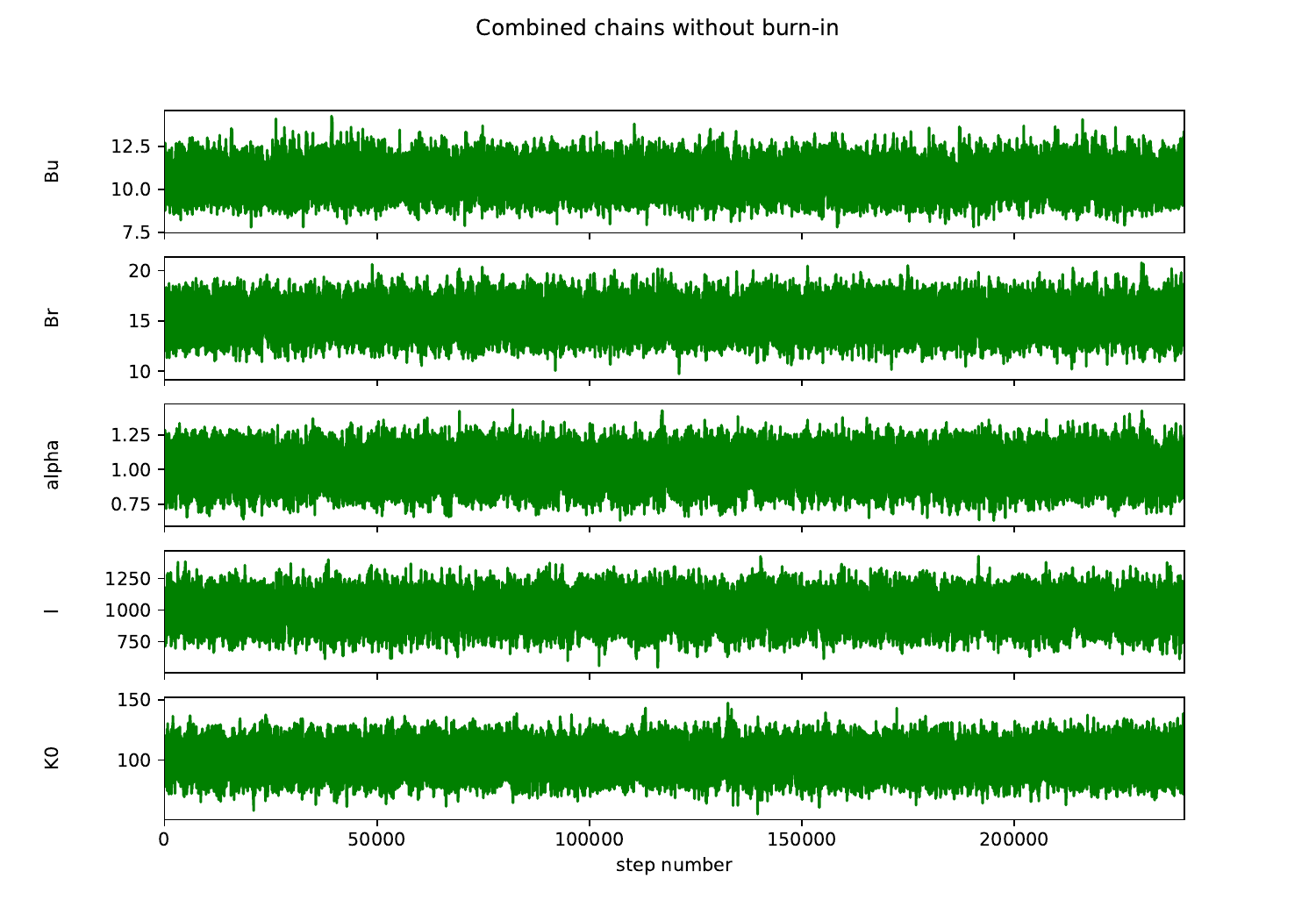}
\caption{Left: Trace plot of a single chain for the parameters from the A-I sampler. Right: Trace plot for 8 merged chains from the A-I method and removed burn-in steps.}
\label{fig:em_traceplot}
\end{center}
\end{figure}

\begin{figure}[h]
\begin{center}
\includegraphics[width=18cm]{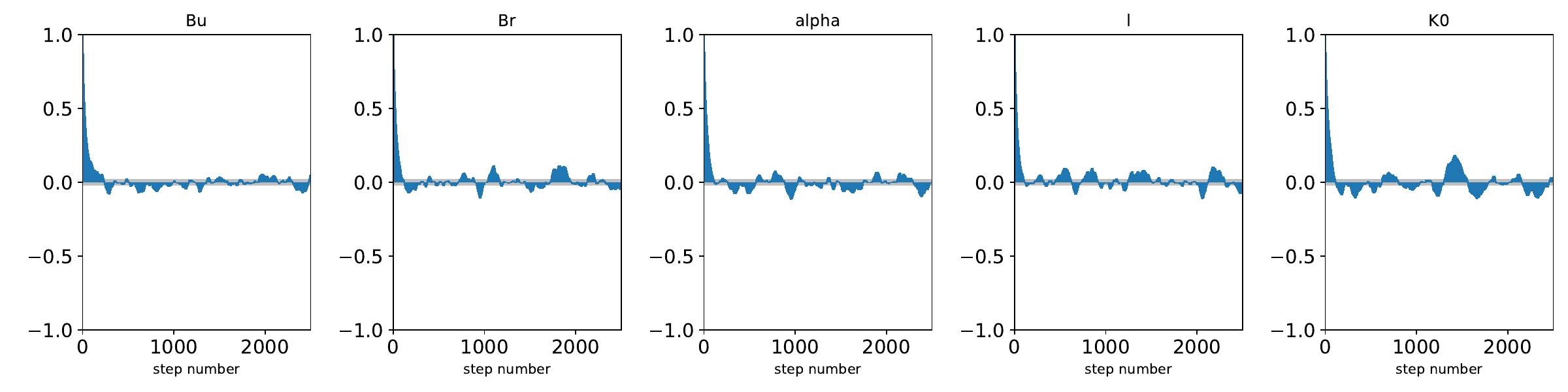}
\caption{Autocorrelation plot for single chain parameters from the MCMC A-I method.}
\label{fig:em_correl}
\end{center}
\end{figure}

\begin{figure}[ht]
    \centering
    \includegraphics[width=0.32\textwidth]{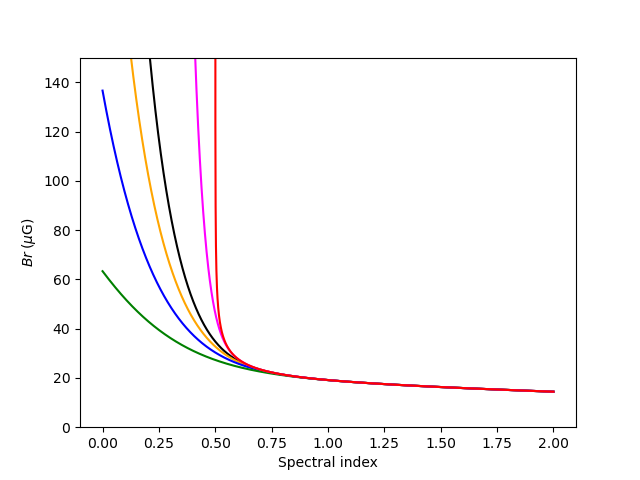}
    \includegraphics[width=0.32\textwidth]{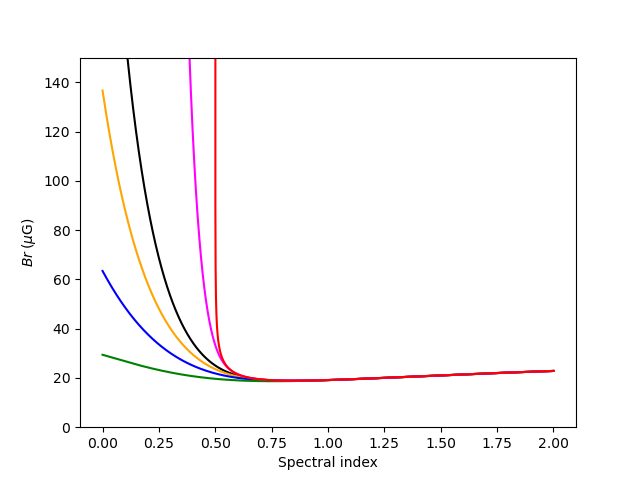}
    \includegraphics[width=0.32\textwidth]{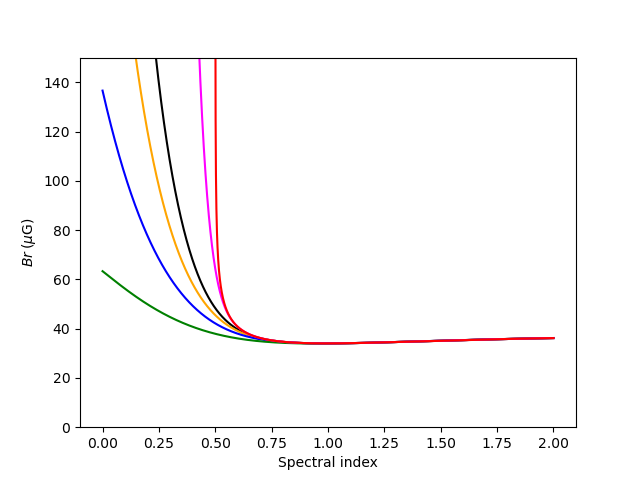}
    \caption{Same as in Figure~\ref{fig:spectral_index_dependence} (left panel) but for 10 times stronger synchrotron intensity (left), 10 times higher frequency (middle), and both changes applied (right). An even larger increase in the frequency of observations (to tens of GHz) would lead to an even steeper increase in the slope of the dependence of $B_r$ on large spectral indices.
    }
    \label{fig:spectral_index_stronger}
\end{figure}

\begin{figure}[ht]
    \centering
    \includegraphics[width=0.32\textwidth]{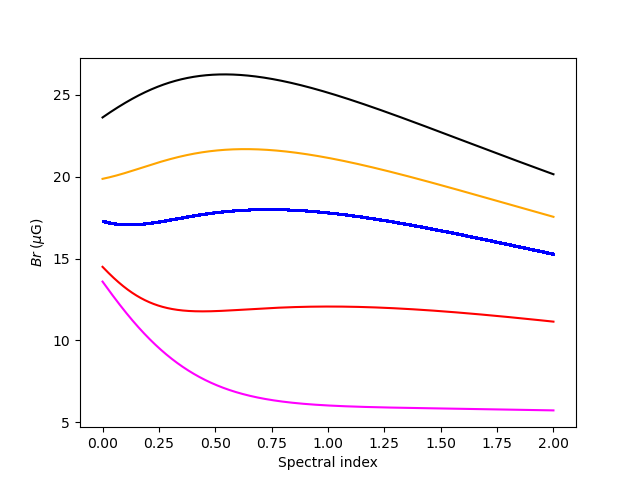}
    \includegraphics[width=0.32\textwidth]{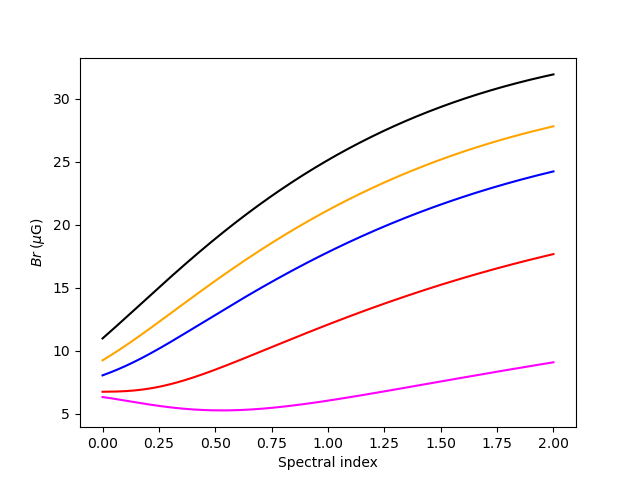}
    \includegraphics[width=0.32\textwidth]{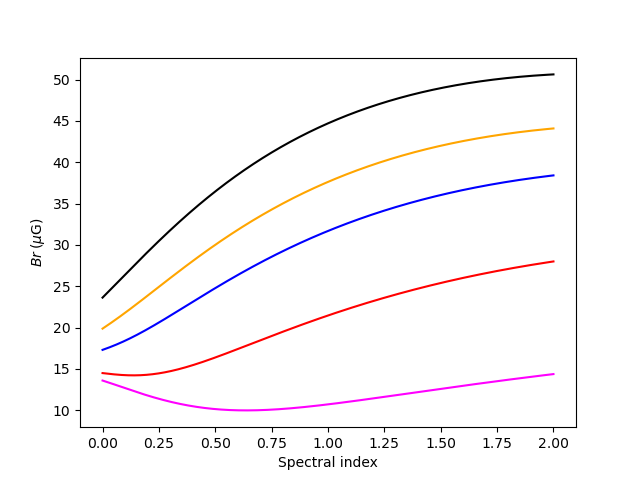}
    \caption{Same as in Figure~\ref{fig:diff_gamma} (left panel) but for 10 times stronger synchrotron intensity (left), 10 times higher frequency (middle), and both changes applied (right). 
    }
    \label{fig:gamma_stronger}
\end{figure}

\clearpage
\bibliography{article}{}
\bibliographystyle{aasjournal}


\end{document}